\documentclass[11pt]{article}
\usepackage{jcapmod}
\usepackage{epsfig}
\usepackage{amsmath, amsfonts,euscript,bbm, amsthm}
\usepackage{ifthen}
\usepackage{graphicx}
\usepackage{color}
\usepackage{bbold}
\usepackage{subfigure}

\newcommand{\e}{\epsilon}
\newcommand{\ssq}{\sigma^{2}_{q}}
\newcommand{\ssi}{\sigma^{2}_{\infty}}
\newcommand{\be}{\begin{eqnarray} }
\newcommand{\ee}{\end{eqnarray} }

\newcommand{\ex}[1]{\langle #1 \rangle}

\newtheorem{thm}{Theorem}
\newtheorem{lem}{Lemma}

\begin{document}

\begin{titlepage}

\setcounter{page}{1} \baselineskip=15.5pt \thispagestyle{empty}

\bigskip\

\vspace{1cm}
\begin{center}

{\fontsize{20}{28}\selectfont  \sffamily \bfseries  Divergence of Perturbation Theory\\[10pt] in Large Scale Structures}


\end{center}

\vspace{0.2cm}

\begin{center}
{\fontsize{13}{30}\selectfont Enrico Pajer and Drian van der Woude}
\end{center}

\begin{center}

\textsl{Institute for Theoretical Physics and Center for Extreme Matter and Emergent Phenomena\\
Utrecht University, Princetonplein 5, 3584 CC Utrecht, The Netherlands}
\vskip 7pt

\end{center}

\vspace{1.2cm}
\hrule \vspace{0.3cm}
\noindent {\sffamily \bfseries Abstract} \\[0.1cm]
We make progress towards an analytical understanding of the regime of validity of perturbation theory for large scale structures and the nature of some non-perturbative corrections. We restrict ourselves to 1D gravitational collapse, for which exact solutions before shell crossing are known. We review the convergence of perturbation theory for the power spectrum, recently proven by McQuinn and White \cite{McQuinn:2015tva}, and extend it to non-Gaussian initial conditions and the bispectrum. In contrast, we prove that perturbation theory diverges for the real space two-point correlation function and for the probability density function (PDF) of the density averaged in cells and all the cumulants derived from it. We attribute these divergences to the statistical averaging intrinsic to cosmological observables, which, even on very large and ``perturbative'' scales, gives non-vanishing weight to all extreme fluctuations. Finally, we discuss some general properties of non-perturbative effects in real space and Fourier space.
\vskip 10pt
\hrule

\vspace{0.6cm}
 \end{titlepage}

\tableofcontents   
		
\newpage


\section{Introduction} 
Exact results in physics are few and far between. Perturbation theory is often the only analytical tool available for both qualitative understanding and quantitative predictions. The dynamics of Large Scale Structures (LSS) is no exception. Because perturbation theory is at the core of most analytic LSS predictions \cite{Bernardeau:2001qr}, it is essential to thoroughly understand its regime of validity and have accurate quantitative estimates of its eventual departure from the exact result. In this work, we make progress in this direction by highlighting the divergence of perturbation theory for LSS correlators in some highly symmetric configurations.

The divergence of perturbation theory in Quantum Field Theory is very familiar to high energy theorist. Already more than 60 years ago, in less than two pages and with only two equations, Dyson showed that perturbation theory for Quantum Electrodynamics (QED) cannot converge and is at best asymptotic \cite{Dyson:1952tj}. His very elegant argument\footnote{The title of this work is an homage to Dyson's classic contribution.} goes as follows. Physical quantities must be analytic functions of the QED coupling constant $\alpha$. Perturbative approximations are given by a power series in $\alpha$. This series must converge in the complex $\alpha$ plane within a ball of radius the distance to the closest singularity. For arbitrarily small but negative $\alpha$ we expect the vacuum to be unstable towards the quantum creation of a large number of pairs of oppositely charged particles. All equally charged particles can be bunched together reducing the energy (since $\alpha<0$) and hence satisfying energy conservation. So any $\alpha<0$ predicts an unstable ground state, which is infinitely different from the free, $\alpha=0$ theory. The radius of convergence is therefore vanishing. 

This elegant argument is intrinsically quantum mechanical in that it relies on pair creation out of the vacuum (violating instantaneous energy conservation in accordance with Heisenberg's uncertainty principle). Yet, the non-convergence of perturbation theory is much more general and ubiquitous even in classical systems with $\hbar=0$. Intuitively, perturbation theory fails to converge whenever the perturbative solutions of the deterministic equations have finite radius of convergence \textit{and} some averaging process, such as the QFT path integral or some stochastic average, probes solutions that lie outside that convergence region. To develop some intuition, let us consider the following toy model. Assume $\delta$ is some random variable (not a space-dependent field). Let us assume that $  \delta $ is related to a Gaussian variable $  \delta_{L} $ in a non-linear way. For concreteness and relevance to LSS studies, we take this relation to be the solution of 1D gravitational collapse\footnote{The time dependence $  \delta(t) $ can be easily added by using $  \delta_{L}(t)=t^{2/3}\delta_{p} $ for some constant $  \delta_{p} $. Since it is irrelevant for this argument we simply drop it.}:
\begin{align}\label{cellsoln2}
\delta=\frac{\lambda\delta_{L}}{1-\lambda\delta_{L}}\,,
\end{align}
where, to emphasize that the perturbation expansion in small $ \delta_{L} $, we introduced\footnote{This can be thought of as the coupling constant of all non-linear terms in the fluid equations \eqref{Euler}, which are all quadratic in perturbations.} a dummy ``coupling constant'' $  \lambda $. We will compute the variance of $  \delta $. The discussion for other cumulants is analogous. By its definition\footnote{The vacuum average $ \ex{1} $ is identically 1 both exactly and in perturbation theory, since $ \lambda $ does not appear in the usual way of computing the integral.}
\be\label{var}
\ex{\delta^{2}}_{\lambda}&=&\int_{-\infty}^{\infty}\frac{d\delta_{L}}{\sigma_{L}\sqrt{2\pi}}e^{-\frac{\delta_{L}^{2}}{2\sigma_{L}^{2}}} \delta^{2}\nonumber \\
&=&\int_{-\infty}^{\infty}\frac{d\delta_{L}}{\sigma_{L}\sqrt{2\pi}}e^{-\frac{\delta_{L}^{2}}{2\sigma_{L}^{2}}} \left(\frac{\lambda\delta_{L}}{1-\lambda\delta_{L}}\right)^{2}\,,
\ee
where we introduced the variance $ \sigma_{L}^{2}$ of the Gaussian random variable $ \delta_{L} $. In more realistic examples, $ \sigma_{L} $ is a function of scale, but for the moment we neglect this complication, i.e. we have zero spatial dimensions.

The integral \eqref{var} does not converge because of the divergence at $ \lambda \delta_{L}=1 $. This pathology is peculiar to the zero-dimensional case and does not play a role in more realistic cases such as 1D and 3D dynamics. Physically, we expect some high density physics to which perturbation theory is not sensitive to prevent the integral from diverging (such as, e.g., pressure). Mathematically, one such example is given by the following regularization
\be\label{reg}
\frac{\delta_{L}}{1-\lambda \delta_{L}}\rightarrow  \frac{1}{\e}\,\arctan\left[  \frac{\e \delta_{L}}{1-\lambda \delta_{L}}\right]\,,
\ee
for some small but finite $  \epsilon $. For any finite $ \e>0 $, the variance $ \ex{\delta^{2}}_{\e\lambda} $ is finite. For small $  \epsilon $, the perturbative expansion of $ \delta $ around $ \lambda=0 $ is independent of $ \e $ and so it is the same as for $  \epsilon=0 $ (up to $  \mathcal{O}(\e) $ corrections). We therefore neglect $  \mathcal{O}(\e) $ corrections in the perturbative expansion in $ \lambda$, whilst keeping in mind that the full, physical result is finite. 

Each order in perturbation theory around $  \lambda=0 $ is finite and the series has the factorial growth typical of asymptotic series\footnote{Here we use the adjective asymptotic to refer to non-convergent asymptotic series, as it is often done in the physics literature, even though of course convergent series are also asymptotic.}. Using
\be\label{poly}
\left(  \frac{\delta_{L}}{1-\lambda \delta_{L}}\right)^{2}&=&\delta_{L}^{2}\sum_{n}^{\infty}(1+n)\left(  \lambda\delta_{L}\right)^{n}\,,\\
\ex{\delta_{L}^{2m}}&=&\left(  \frac{\sigma_{L}^{2}}{2}\right)^{m}\frac{(2m)!}{m!}\,,
\ee
we find
\be\label{series}
\ex{\delta^{2}}_{\lambda}^{\mathrm{PT}}&=&\sum_{m=0}^{\infty}(1+2m)\left(  \frac{\lambda\sigma_{L}}{\sqrt{2}}\right)^{2m}\frac{\sigma_{L}^{2}}{2}\frac{\left(  2m+2\right)!}{(m+1)!}\,.
\ee
One can the use Stirling formula to expand this for large $ m $
\be
\ex{\delta^{2}}_{\lambda}^{\mathrm{PT}, m}&\sim&\,4\sqrt{2}m^{2}\sigma_{L}^{2}\left(  \frac{2m\lambda^{2}\sigma_{L}^{2}}{e}\right)^{m}\,.
\ee
The perturbative calculation therefore starts diverging at order $ n=2m\simeq e/(\lambda^{2} \sigma_{L}^{2}) $ and therefore 
\be
\ex{\delta^{2}}_{\lambda}^{\mathrm{PT}}&\neq&\ex{\delta^{2}}_{\lambda}\,.
\ee 
Summarizing, we have proven that the perturbative series \eqref{series} does not converge for any finite value of $ \lambda $, it has zero radius of convergence. The series is nevertheless asymptotic to the right (regularized) answer because at every finite perturbative order $  n $ one has
\be
\lim_{\lambda\rightarrow0}\left[ \ex{\delta^{2}}_{\lambda}^{\mathrm{PT,n}}-\ex{\delta^{2}}_{\lambda} \right]=0\,.
\ee
What happened? The non-linear relation \eqref{cellsoln2} between $  \delta $ and $  \delta_{L} $ admits a perturbative approximation around $  \lambda=0  $ (equivalently $  \delta_{L}=0 $, but for extra clarity we formulate it in terms of the fictitious coupling constant) that has a finite radius of convergence $  |\lambda \delta_{L}|<1 $. But the average in \eqref{var} extends all the way to $  \delta_{L}=\infty $. The result is that every perturbative correction to the variance contains an error coming from the exponentially damped tails of the integral. This error grows with the perturbative order because of the growth of the order of the polynomial approximation in \eqref{poly}. In more realistic cases such as 1D or 3D dynamics, we believe it is still true that the perturbative solution of the deterministic equations of motion has finite radius of convergence. The exact solutions to planar and spherical collapse we discuss in section \ref{sec:2} support this idea. One therefore generically expects perturbation theory to be divergent (and asymptotic) also in more realistic cases. In this work, we show explicitly that this is the case for real space correlators and count-in-cell statistics in 1D. Perhaps surprisingly, but as anticipated in \cite{McQuinn:2015tva}, perturbation theory instead converges in Fourier space. We are certainly not the first to investigate the convergence of perturbation theory for LSS \cite{McQuinn:2015tva,Blas:2013aba,Afshordi:2006ch,Valageas:2013hxa,Mohammed:2014lja,Seljak:2015rea,Valageas:2007ge,Valageas:2009km,Tassev:2012cq,Valageas:2010rx,Valageas:2001df,Baldauf:2015tla,Tatekawa:2006gx}, and we refer to the relevant literature in due course. 

It is important to stress that the non-convergence of perturbation theory we discuss in this work has nothing to do with the improvements advocated by the EFT of LSS \cite{Baumann:2010tm}. This is easily seen since the EFT corrections arise from smoothing short scale dynamics and hence disappear as we take the short scale power to zero. The non-convergence we discuss here instead does not disappear in this limit. More intuitively, the EFT of LSS corrections captures the effect of the non-perturbative short scales on large scales. Non-convergence of perturbation theory instead results exclusively from large scales, with arbitrarily small power. We come back to this point in section \ref{EFT}.

There are actually two conceptually distinct ways in which perturbation theory can fail to approximate some desired result: it might not converge, as we have just seen, or perturbation theory might converge to a result that is not the right one. The second situation arises also in all realistic LSS computation. In both Eulerian and Lagrangian approaches one cannot fully capture multistreaming (but see \cite{McDonald:2017ths} for an exception) and therefore even if perturbation theory converged, it would not describe the correct physical result. While we originally attempted to makes progress in this direction as well, we have been able only to derive rough estimates for the non-perturbative corrections coming from multi-streaming. We have collected them with some general remarks in section \ref{sec:5}.

Before diving into the derivation of our results, it is important to explain why one should care about non-perturbative results, since their exponentially small amplitude is typically trumped by larger perturbative corrections. There are several reasons. First, since the dawn of time, analytical approaches to LSS have been trying to push predictions closer and closer to the non-linear scale, where all perturbative approximations break down. Around the non-linear scale perturbative and non-perturbative corrections both become of order one! Which one is largest might depend on numerical factors that are impossible to predict a priori. Therefore, a conservative estimate of the theoretical error of perturbative methods, as advocated recently in \cite{Baldauf:2016sjb,Welling:2016dng}, should include non-perturbative corrections as well. Second, some non-perturbative corrections might break symmetries that are respected by perturbative terms. Tunneling in quantum mechanics for example is invisible to perturbation theory (see e.g. \cite{Marino:2015yie}). In the context of LSS, scale dependent bias \cite{Dalal:2007cu} is a relevant example: it is a non-perturbative effect (since the tracers of interest are non-perturbative objects) that cannot be mimicked by standard, late-time gravitational evolution as consequence of the equivalence principle. It would be nice to find other non-perturbative observables with an equivalent sensitivity to primordial initial conditions. Last but not least, a physicist has so few occasions to glimpse at what lies beyond perturbation theory that any chance should be taken advantage of.

For the convenience of the reader we summarize here our main results.
\begin{itemize}
\item The convergence of 1D Standard Perturbation Theory (SPT) to the Zel'dovich result (ZA), which is exact in 1D before shell crossing, was recently established by McQuinn and White \cite{McQuinn:2015tva} for $  \Lambda $CDM-like initial conditions. We review their derivation and formalize one technical but crucial step. We stress that convergence relies on the (realistic) assumption that the variance of the displacement is finite. In fact, SPT is shown to diverge for scaling universes with a negative spectral tilt $  P_{L}\propto k^{n} $ with $  -1<n\leq 0 $ \cite{Simon} (see subsection \eqref{scalingpk}). We generalize the convergence result for $  \Lambda $CDM-like initial conditions to the bispectrum and for non-Gaussian initial conditions.
\item We prove analytically and verify numerically that perturbation theory does \textit{not} converge instead for the real space equivalent, namely the correlation function. The technical reason is that the Fourier transform integral cannot be interchanged with the infinite sum over perturbative contributions. More intuitively, we show that the reason for non-convergence is a non-perturbative tail contribution to the correlation function, similar to the toy model above.
\item The relevance and potential non-perturbativity of tails of the probability distribution function for the average density $\delta_{R}$ in a cell of radius $R$ has been noted in various places, e.g. \cite{Valageas:2009km,Bernardeau:2013dua}. We prove and verify numerically that, in the context of their 1D equivalent, there is indeed a finite radius of convergence for perturbation theory for this PDF. We show that any perturbative computation of cumulants is therefore asymptotic. Again, we highlight the analogy with the toy model. 
\item Along the way, we present a new derivation of this count-in-cell PDF in 1D, which is unitary with unit mean by construction.
\end{itemize}

The rest of this paper is organized as follows. In section \ref{sec:2}, we collect standard results about exact solutions for gravitational collapse. We discuss the radius of convergence of perturbation theory show how nonlinear transformations can improve convergence. Section \ref{sec:3} contains the main results about the convergence of perturbation theory for Fourier space correlators and the non-convergence for real space correlators. Section \ref{sec:4} is dedicated to the construction of the 1D count-in-cell PDF, with details in Appendix \ref{PDF}, and a proof of the finite radius of convergence of perturbation theory. In section \ref{sec:5} we draw some qualitative conclusions about the existence and relevance of non-perturbative effects in real and Fourier space. We conclude in Section \ref{sec:6}.  


\section{Exact and perturbative classical solutions to gravitational collapse}\label{sec:2}

In this section we discuss perturbation theory of the classical equations of motion for LSS and its convergence properties. This discussion is logically separated from the discussion of statistical/quantum correlators, which we postpone to the following sections. In the following, we review some well-known exact solutions to gravitational collapse in 1D. We follow mostly the review part of \cite{McQuinn:2015tva}.  One can think of 1D gravitational collapse as a more symmetric version of 3D collapse, in which the density field is only allowed to vary in one direction, say along $x$. It is thus the problem of the evolution of 2D-homogeneous and isotropic sheets of matter, with density contrast
\begin{align}
\delta(x)=\frac{\rho(x)}{\bar{\rho}}-1,
\end{align}
where $x$ is just a number in this case and we omitted the time dependence. Moreover, we restrict to an Einstein-de Sitter spacetime background, for which
\begin{align}\label{EdS}
\bar{\rho}(a)=\bar{\rho}(a_{i})\left(\frac{a_{i}}{a}\right)^{3}.
\end{align}
The equations of motion are obtained by imposing the symmetries of the fluid equations that are assumed to hold in Standard Perturbation Theory (SPT) (for a review see \cite{Bernardeau:2001qr}). Except in section \ref{scalingpk}, we neglect Effective Field Theory (EFT) corrections \cite{Baumann:2010tm}, since introducing these terms should not change our results qualitatively, but considerably complicates the algebraic manipulations. To consistently neglect them, we exponentially damp the initial power spectrum, such that all fields can be thought of as smoothed fields. The equations of motion are then
\begin{align}\label{Euler}
\partial_{\tau}\delta+\theta &=-\nabla(\delta v)\,, \nonumber \\
\partial_{\tau}\theta+\mathcal{H}\theta+4\pi G a^{2}\bar{\rho}\delta &=-\nabla(v\nabla v),
\end{align}
where $v$ is the velocity field, $\theta=\nabla v$, and $\tau$ and $\mathcal{H}$ are the conformal counterparts of the time coordinate and the Hubble rate, respectively. Here we have taken the gradient of the Euler equation without loss of generality, as in 1D there are no vector modes, and we used the Poisson equation to get rid of the Newtonian potential $  \phi $: 
\begin{align}\label{Poisson}
\Delta \phi = 4\pi G a^{2}\bar{\rho}\delta.
\end{align}

As we review below, the Zel'dovich approximation is the exact solution to 1D gravitational collapse before shell-crossing, and its implications for the density field are straightforward. We show that the same solution holds for the evolution of the average density in cylindrical cells, which are effectively 1-dimensional cells. This analysis is very similar to spherical collapse. Given these exact solutions, we investigate the convergence of the perturbative solutions. We show that the perturbative solutions in real space have a finite radius of convergence. We comment on how non-linear transformations can provide non-perturbative improvements in the convergence.     

\subsection{Zel'dovich solution}

The Zel'dovich solution \cite{Zeldovich:1969sb} is exact before shell crossing in 1D \cite{Schandarin:1989sr,Sahni:1995rm}. This follows from the fact that in 1D Newtonian Gravity, force is independent of distance. The equation for the gradient of the displacement field, giving the displacement of a fluid element from its initial position $q$, turns out to be linear
\begin{align}\label{eqpsi}
\nabla_{q}\left[\Psi^{\prime\prime}(q)+\mathcal{H}\Psi^{\prime}(q)\right]=4\pi G a^{2}\bar{\rho}\nabla_{q}\Psi,
\end{align} 
where primes denote derivatives with respect to conformal time, and we used
\begin{align}
1+\delta(x)=\int dq \delta_{D}[x-q-\Psi(q)]=\frac{1}{1+\nabla_{q}\Psi}\bigg|_{x=q+\Psi(q)},
\end{align}
which in Fourier space reads
\begin{align}\label{ZAF}
\delta(k)=\int dq \,e^{-ikq}\left(e^{-ik\Psi(q)}-1\right).
\end{align}
One can check that this definition of $\delta(x)$ indeed yields a solution to the Euler-Poisson system \eqref{Euler}, \eqref{Poisson} for $\Psi(q,a)=a/a_{i}\Psi(q,a_{i})$, which solves \eqref{eqpsi}. In particular, upon the identification 
\begin{align}
\left(\partial_{\tau} |_{x}+v\partial_{x}\right)=\partial_{\tau} |_{q} \, ,
\end{align} 
where we indicated what is kept fixed when performing the time-derivative, $\delta(q)$ satisfies the Lagrangian equation
\begin{align}\label{Leom}
\delta^{\prime\prime}(q)+\mathcal{H}\delta^{\prime}(q)-2\frac{\delta^{\prime 2}(q)}{1+\delta(q)}=4\pi G\bar{\rho}\delta(q)(1+\delta(q)).
\end{align}
Introducing the linear order density $\delta_{L}(q,\tau)\equiv -\nabla_{q}\Psi(q,\tau)$, we write
\begin{align}\label{1Dsolution}
1+\delta(x,\tau)=\frac{1}{1-\delta_{L}(q,\tau)}.
\end{align}
A few comments are in order. First, observe that the analytic properties for the solution for $\delta$ in real space seem different from those in Fourier space. We come back to this issue below. Second, the Lagrangian equation \eqref{Leom} is actually also found for the evolution of the density in finite cells. We derive this in Appendix \ref{planar} within Newtonian cosmology. Third, as expected, \eqref{1Dsolution} breaks down for overdensities when the density blows up, which is precisely when shell-crossing occurs. On the other hand, this solution is well defined at all times for underdensities, whose density asymptotes to $-1$.


\subsubsection{Perturbative solution}

A more elaborate analysis of SPT for the 1D Euler-Poisson system \eqref{Euler}, \eqref{Poisson} was done in \cite{McQuinn:2015tva}, in which they showed that the Fourier kernels \cite{Grinstein:1987qf} obtained from SPT are equivalent to the ones obtained by expanding the ZA solution \eqref{ZAF}. Here we are more modest, and just consider the `Lagrangian' problem of the evolution of the density in a fluid element or cell \eqref{Leom}. Formally, one can solve \eqref{Leom} (or its cosmological time equivalent, \eqref{1Dcollapse}) perturbatively, using a Green's function method
\begin{align}
\delta=\delta_{L}+\int dt^{\prime}G(t,t^{\prime})\left[2\frac{\dot{\delta}^{2}}{1+\delta}+4\pi G\bar{\rho}\delta^{2}\right],
\end{align} 
where we have selected the growing mode linear solution, and 
\begin{align}
D_{t}G(t,t^{\prime})= \delta_{D}(t-t^{\prime}),
\end{align}
for the linear differential operator $D_{t}$ in \eqref{1Dcollapse}. The perturbative solution is then obtained by iteratively plugging the lower order solutions into the nonlinear terms. In an Einstein-de Sitter universe, this leads to a power series in $\delta_{L}$. However, since we already know the full solution \eqref{1Dsolution}, this has to coincide with a simple expansion of the exact solution in $\delta_{L}$, leading to the following perturbative solution:
\begin{align}\label{1DPT}
\delta_{PT}^{(n)}=\sum_{i=1}^{n}c_{i}\delta_{L}^{i},
\end{align}
for some constants $c_{i}$, and $c_{1}=1$.

\subsubsection{Convergence}

Since the exact solution \eqref{1Dsolution} has a nice analytic form, we can directly apply standard results from complex analysis about the convergence properties of the perturbative series \footnote{This is an example of the idea put forth in footnote 5 of \cite{Porto:2013qua}.}. In particular, the radius of convergence around the origin ($\delta_{L}=0$) is given by the distance to the nearest pole, which is in our case is $\delta_{L}=1$. For overdensities, this makes sense, as this is the point beyond which the exact solution breaks down as well. In other words, the series converges for overdensities all the way to $  \delta=+\infty $. However, this radius of convergence also implies that, for underdensities, the series only converges up to $\delta_{L}=-1$, for which $\delta=-1/2$, whereas the exact solution sensibly extends all the way to $\delta=-1$. 

\begin{figure}[h]
	\centering
	\includegraphics[width=\columnwidth]{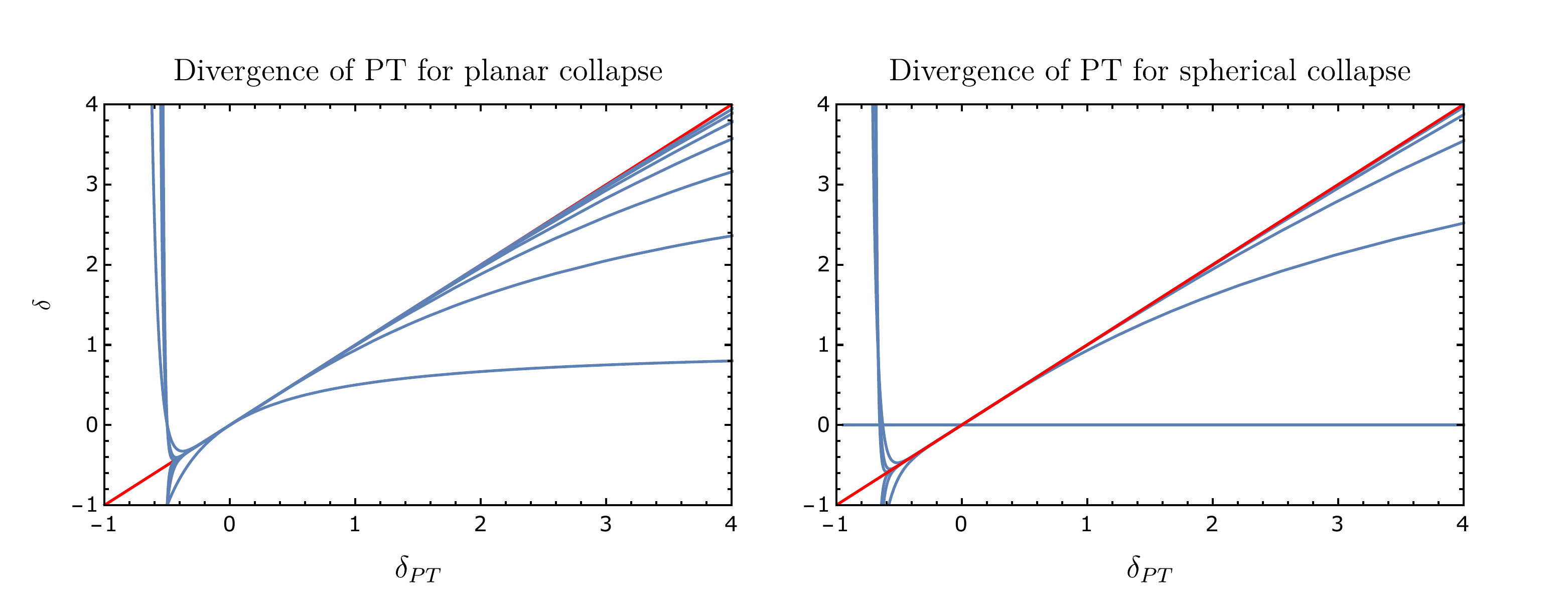}
	\caption{The parametric plots shows $\delta$ versus $\delta_{PT}$. The blue lines show $\delta_{PT}$ from linear to 18-loop order in steps of three loops. The red reference line is the simple diagonal $\{\delta,\delta\}$. These plots show the divergence of perturbation theory beyond $\delta<-0.5$ for planar collapse \eqref{1Dsolution} (left) and beyond $\delta<-0.684$ for spherical collapse \eqref{deltaSC} (right).\label{PTvsExact}}
\end{figure}

One way to visualize the performance of perturbation theory is to plot $\delta_{PT}^{(n)}$ against the exact solution for all times. So we plot the following points 
\begin{align}
\left\{\left(\delta(a),\delta_{PT}^{(n)}(a)\right) \| a\in \left[-\infty,\infty\right]\right\}
\end{align}
in Figure \ref{PTvsExact}. The non-convergence beyond $\delta=-1/2$ is clearly visible. 

\subsubsection{Nonlinear transformations and improved convergence}

As noted above\footnote{Results in this subsection were obtained in collaboration with Gabriele Trevisan.}, \textit{the radius of convergence of the perturbative expression is smaller than the radius for which the densities are physically well defined, which is the reason perturbation theory does not converge for all physically relevant densities.} One could ask if nonlinear transformations could fix this problem. Here we show that the answer is yes. One option is to choose an invertible, analytic function over the whole real axis, whose range is at least $(-1,\infty)$. Trivial examples are, on top of the linearizing transformation \eqref{1Dsolution}, 
\begin{align}
\delta(a)=e^{\lambda(a)}-1 \quad \text{or}\quad \delta(a)=2e^{\lambda(a)}-2. 
\end{align} 
This corresponds to a nonlinear transformation of the perturbation parameter of the form $\lambda=-\log (1-\delta_{L})$. Perturbation theory in $  \lambda $ now converges for all physically meaningful values, namely in the whole interval $\delta\in \{-1,\infty\}$.

Two comments are in order. First, this example shows that not all perturbation schemes are equivalent. In particular there can very well be non-linear transformations that substantially improve the convergence of perturbation theory. Second, the improvement of convergence above was only possible because we knew the full result and could therefore guess the correct non-linear transformation. It is not clear whether in more complicated cases, such as the full 3D dynamics, this can be achieved. For empirical attempts in this direction see \cite{Neyrinck:2009fs,Neyrinck:2010th,Coles:1991if,Wang:2011fj,Carron:2011et,Hall:2017pzf} for logarithmic and Gausianizing transformations, and \cite{Simpson:2011vn,Simpson:2013nja} for clipping procedures, in which large overdensities are taken out of the ensemble averages. 

\subsection{Spherical collapse}

Let us now consider the collapse of a spherical overdensity in a spherically symmetric universe (see, e.g. \cite{Mukhanov:2005sc,Peebles:1980}). The density contrast is a function of radius $R$ only:
\begin{align}
\delta(R)=\frac{\rho(R)}{\bar{\rho}}-1.
\end{align} 
Throughout, we assume the background density is the EdS one, \eqref{EdS}. We are interested in the evolution of the density inside a spherical cell. The total mass inside the cell is
\begin{align}\label{Mass3D}
M=4\pi \int_{0}^{R}dr r^{2}\rho(r).
\end{align} 
Then, by spherical symmetry, Gauss' law yields the following flux perpendicular to the surface,
\begin{align}
4\pi G M=4\pi R^{2}\nabla_{r}\phi,
\end{align} 
leading of course to the spherical collapse equation
\begin{align}\label{Scollapse}
\ddot{R}=-\frac{GM}{R^{2}}.
\end{align}
Rewriting equation \eqref{Mass3D} in terms of the average density contrast,
\begin{align}
\delta_{R}=\frac{3}{ R^{3}}\int_{0}^{R} dr r^{2} \delta(r),
\end{align}
we can express $R$ in terms of the average density
\begin{align}\label{deltaradius}
\delta_{R}=\frac{3M}{4\pi R^{3}\bar{\rho}}-1.
\end{align}
Plugging this into \eqref{Scollapse}, we obtain the evolution equation for the average density in this spherical cell:
\begin{align}\label{3Dcollapse}
\ddot{\delta}_{R}+2H\dot{\delta}_{R}-\frac{4}{3}\frac{\dot{\delta}_{R}^{2}}{1+\delta_{R}}=4\pi G\bar{\rho}\delta_{R}(1+\delta_{R}).
\end{align}
Once again, one finds the same equation for infinitesimal volume elements in Lagrangian coordinates from the fluid equations in a spherically symmetric setup. Note that this equation, as the planar one, only depends on the density, there is no explicit mention of mass or scale. This is a consequence of the fact that the spherical collapse equation is symmetric under rescalings that leave $M/R^{3}$ - the density - fixed. Similarly, for planar collapse, rescalings that leave $M/R$ - the 1D density, up to the Hubble expansion in the orthogonal directions - fixed, are a symmetry.

\subsubsection{Exact solution}

Despite its similarity to the planar case, the solution to \eqref{3Dcollapse} is only known in parametric form. Moreover, depending on whether the initial density perturbation is positive or negative, the form of the solution is slightly different. For overdensities, one can check that \eqref{Scollapse}, and therefore \eqref{3Dcollapse} are solved by
\begin{align}
R=A(1-\cos\eta); \quad t=B(\eta-\sin\eta)+C, 
\end{align}
provided
\begin{align}
\frac{A^{3}}{B^{2}}=GM,
\end{align}
and $\eta \in [0,2\pi]$, as can be seen from the expression for R. For underdensities, we find
\begin{align}
R=A(\cosh\eta-1); \quad t=B(\sinh\eta-\eta)+C, 
\end{align}
with the same restriction on the constants $A$ and $B$ and $\eta \in [0,\infty]$ this time. The solutions are found by plugging this into \eqref{deltaradius}, which for overdensities becomes
\begin{align}\label{SCexact}
\delta=\frac{3M}{4\pi A^{3}(1-\cos\eta)^{3}\bar{\rho}}-1.
\end{align}
Note that this expression requires the time dependence of $\bar{\rho}$, which in an EdS universe is given by
\begin{align}
\bar{\rho}=3M_{pl}^{2}H^{2}=\frac{1}{6\pi G t^{2}}.
\end{align}
Hence,
\begin{align}\label{deltaSC}
\delta=\frac{9}{2}\frac{GMt^{2}}{A^{3}(1-\cos\eta)^{3}}-1=\frac{9}{2}\frac{t^{2}}{B^{2}(1-\cos\eta)^{3}}-1.
\end{align}
To find $\delta(t)$, we need to invert the relation between $t$ and $\eta$. This gives the solution for $\delta$ as a function of two constants, $B$ and $C$, as it should for a second order differential equation. For a more familiar interpretation of these constants in terms of the growing and decaying mode, we need to restrict ourselves to the small density regime. As we show in appendix \ref{app:12}, this is given by
\begin{align}
\delta_{L}=\frac{3}{10}\left(\frac{9}{2}\right)^{1/3}\left(\frac{t}{B}\right)^{2/3}+\frac{2C}{t}.
\end{align}
These are indeed the familiar growing and decaying modes, parametrized by $B$ and $C$ respectively. This makes manifest that for adiabatic initial conditions, we should set $C=0$. At the same time, this clarifies the range of validity of the growing mode solution. The initial conditions are set by
\begin{align}
\delta_{i}=\frac{3}{10}\left(\frac{9}{2}\right)^{1/3}\left(\frac{t_{i}}{B}\right)^{2/3}.
\end{align}
As argued before, the solution for overdensities only makes sense up to $\eta=2\pi$ - the point of collapse. This means 
\begin{align}
2\pi=\frac{t}{B}=\left(\frac{3}{10}\right)^{3/2}\left(\frac{9}{2}\right)^{1/2}\frac{t}{t_{i}}\delta_{i}^{3/2}.
\end{align}
In other words, the solution is well defined up to the present for initial conditions that satisfy
\begin{align}
\delta_{i}<\delta_{c}\left(\frac{t_{i}}{t_{0}}\right)^{2/3}=\delta_{c}\frac{a_{i}}{a_{0}},
\end{align}
where
\begin{align}
\delta_{c}=\frac{10}{3}\left(\frac{9}{2}\right)^{1/3}\left(2\pi\right)^{2/3}\approx 1.686.
\end{align}
One can check that, up to a minus sign, the growing and decaying mode are the same for underdensities. In the underdense case, the fully non-linear, growing mode solution is well defined for all initial conditions (larger than -1) and all times.


\subsubsection{Perturbative solution and convergence}\label{Gab1}

The perturbative solution to the equation of motion still leads to a series expansion in $\delta_{L}$, which we can obtain  from the exact solution as follows. We are looking for a solution of the form 
\begin{align}\label{SCPT}
\delta(t)=\sum_{i}^{n}c_{i}\delta_{L}^{i}+\mathcal{O}(\delta_{L}^{n+1})=\sum_{i}^{n}\tilde{c}_{i}\left(\frac{t}{B}\right)^{2i/3}+\mathcal{O}\left(\left(\frac{t}{B}\right)^{2/3(n+1)}\right).
\end{align} 
We chose to keep the parameter $B$ explicitly, so that the both the left hand side (see \eqref{deltaSC}) and the right are functions of $\eta$ only. Expanding in $\eta$ and matching order by order then allows us to solve for the $\tilde{c}_{i}$, which are directly related to the $c_{i}$.
 
Once again, we can study the convergence of this series, previously also discussed in \cite{Porto:2013qua,Sahni:1995rr}. Observe that the series breaks down for overdensities when the density blows up, which is precisely when $\delta_{L}=\delta_{c}$. From the Green's function approach, it is clear that the perturbative solution has to be the same for over- and underdensities; the only thing that distinguishes between them is whether $\delta_{L}$ is positive or negative. This can of course be checked explicitly applying the above logic to the underdense solution. Thus, the same rules of complex analysis tell us that the series for underdensities only converges up to the point where $\delta_{L}^{u}=-\delta_{c}$, which by definition corresponds to 
\begin{align}
\frac{t_{c}}{B}=2\pi.
\end{align}
The critical $\eta$ parameter is then found from the relation between $t$ and $\eta$ for underdensities:
\begin{align}
\sinh \eta_{c}-\eta_{c}=\frac{t_{c}}{B} \quad \implies \quad \eta_{c}\approx 2.915.
\end{align}
Plugging this back into the full solution, we find that perturbation theory only converges in the range $ -0.684<\delta<+\infty$. Again, in Figure \ref{PTvsExact} we plot $\delta_{PT}$ versus $\delta$ for spherical collapse to visualize these statements. 


\subsubsection{Improved convergence}\label{Gab2}

Similar to the 1D case, there are ways to improve the radius of convergence of perturbation theory, knowing the full solution\footnote{Results in this subsection were obtained in collaboration with Gabriele Trevisan.}. In this case, a neat example is found from observing that the underdensity solution is obtained from the overdensity solution by rotating in the complex plane $\eta\to i\eta$. This immediately tells us that the perturbative expansion of $\delta$ in terms of $\eta$ converges to $|\eta|=2\pi$, as this is the radius of convergence for the overdense solution. This is a much larger value than $\eta_{c}$ we found above. In fact, the final underdensity at this value of $\eta$ is $\delta=-0.984$. So, once again, a non-linear transformation of the expansion parameter from $  \delta $ to $  \eta $ does enlarge the physical radius of convergence of the theory (to the range $-0.984<\delta<+\infty$). The analogous non-linear transformation in realistic 3D cases can also be searched for heuristically \cite{Neyrinck:2009fs,Neyrinck:2010th,Simpson:2011vn,Simpson:2013nja}.


\section{(non-)Convergence of PT for 1D correlators}\label{sec:3}

In this and the following sections, we move away from the discussion of the classical solutions of the equations of motion and delve into the computation of stochastic/quantum correlators. There are already various studies on the reach of perturbation theory for large scale structures in the literature. The relevance of halos in this context was stressed in \cite{Afshordi:2006ch,Valageas:2013hxa,Mohammed:2014lja,Seljak:2015rea}. The reach of PT was further analyzed in \cite{Valageas:2007ge,Valageas:2009km,Tassev:2012cq}, and the relevance of shell crossing was studied in \cite{Valageas:2010rx,Valageas:2001df,Pueblas:2008uv}. Finally, a generic perturbative expression including EFT corrections for the power spectrum was tested numerically in \cite{Baldauf:2015tla}. In this section we analytically test these ideas by studying the convergence of PT for 1D correlators. Our main finding here is that, for $  \Lambda $CDM-like initial conditions, SPT \cite{Bernardeau:2001qr} converges to the correct power spectrum and bispectrum both for Gaussian and non-Gaussian initial conditions. The convergence of SPT for the power spectrum was to a large extent already established by McQuinn and White in \cite{McQuinn:2015tva}. In subsection \ref{LCDMpk}, we review their derivation and extend it marginally by rigorously justifying their final step, namely that one can safely exchange the integral over initial positions with the infinite perturbative sum. While this seemingly minor technical assumption is justified for the power spectrum of LCDM, it is actually invalid in a few relevant cases. In fact, for scaling universes, $  P(k)=A k^{n} $, SPT diverges for $  -1<n<0 $, while it converges for $  n>0 $ \cite{Simon} (see subsection \ref{scalingpk}). More importantly, the exchange of sum and integral is not allowed for real space correlators and leads to the non-convergence of SPT, a new result which we discuss in the section \ref{nonccf}. 


\subsection{Prerequisites}

We start by collecting the ingredients necessary for the derivations below. The key mathematical observations that lead to our results are explained first. Then we define the Zel'dovich approximation (ZA) for correlators and recall some properties of the initial conditions of our universe that are important for what follows. 

\subsubsection*{Mathematical prerequisites}

We will see that the ZA allows us to write all observable as some integral, of the form
\begin{align}\label{modelintegral}
\hat{O}(x,\sigma^{2})=\int dy\, f(y,x,\sigma^{2}),
\end{align}
where $\sigma^{2}$ is a dimensionless parameter representing the size of the linear power spectrum. We wish to answer the question whether perturbation theory in this parameter resums to the ZA result. This relies on two steps:
\begin{itemize}
	\item Can we write $f$ as a convergent power series in $\sigma^{2}$?
	\item Can we interchange the order of integral and sum?
\end{itemize}

For the convergence proofs of \textit{Fourier space} observables in the remainder of this paper, it turns out that both of these questions can be answered positive. The first step is always obvious. For the second step, we need to invoke the Fubini-Tonelli theorem.\footnote{To show the subtleties of this second step, consider the Fourier transform of a Gaussian:
\begin{align}
\hat{O}(k,\sigma^{2})=\int dq\, e^{i q k} \frac{1}{\sqrt{2\pi}}e^{-\frac{q^{2}\sigma^{2}}{2}}=\frac{1}{\sqrt{2\pi\sigma^{2}}}e^{-\frac{k^{2}}{2\sigma^{2}}}.
\end{align}
Clearly, the final expression for $\hat{O}$ is non-analytic around $\sigma^{2}=0$, whereas the integrand in the middle step is analytic (in fact it can be extended to an entire function). Apparently,
\begin{align}
\int dq\, \sum_{i}f_{i}(q,k)\left(\sigma^{2}\right)^{i}\neq \sum_{i}\int dq\,f_{i}(q,k)\left(\sigma^{2}\right)^{i}, \quad \text{for} \quad  f(q,k,\sigma^{2})=e^{i q k} \frac{1}{\sqrt{2\pi}}e^{-\frac{q^{2}\sigma^{2}}{2}},
\end{align} 
where $f_{i}$ are the series coefficients. In fact, the integrals over the $f_{i}$, which are simple power laws in this case, only make sense as a distribution. The resulting expression on the right hand side is then a sum of derivatives of the Dirac-delta distribution
\begin{align}
\sum_{i}\int dq\,f_{i}(q,k)\left(\sigma^{2}\right)^{i}=\sum_{i}\delta_{D}^{(2i)}(k)c_{i}\sigma^{2i},
\end{align} 
for some $c_{i}$. One can now check that the left and right hand side are \textit{not} equal as a distribution acting on test functions $\varphi$, since
\[
\begin{cases}
\int dk\, \frac{1}{\sqrt{2\pi\sigma^{2}}}e^{-\frac{k^{2}}{2\sigma^{2}}}\varphi(k)=\int dk\, \sum_{i}\delta_{D}^{(2i)}(k)c_{i}\sigma^{2i}\varphi(k) \quad \text{for analytic $\varphi$} \nonumber \\
\int dk\, \frac{1}{\sqrt{2\pi\sigma^{2}}}e^{-\frac{k^{2}}{2\sigma^{2}}}\varphi(k)\neq \int dk\, \sum_{i}\delta_{D}^{(2i)}(k)c_{i}\sigma^{2i}\varphi(k) \quad \text{for non-analytic $\varphi$},
\end{cases}
\]
Thus interchanging sum and integral is not allowed, even if interpret functions as distributions. } It states: 
\begin{thm}\label{F-T}
(Fubini-Tonelli:) If $\int dq \sum_{i} |f_{i}(q)|<\infty$, then $\int dq \sum_{i} f_{i}(q)=\sum_{i}\int dq  f_{i}(q)$.
\end{thm}
Below we show that for the computation of LSS correlators, the theorem can be used, with some minor subtlety, for \textit{Fourier space} observables.

For the \textit{real space correlation function} (as well as the 1D PDF), the procedure is different. In that case we prove that the power series in $\sigma^{2}$ of the integrand in \eqref{modelintegral} is divergent for $|y|$ larger than some fixed value. In order to conclude that this leads to a divergent series for the integral as well, we make use of the following lemma. 
\begin{lem}\label{assumption}
	If 
	\begin{align}\label{ass2}
	\left(\frac{\partial}{\partial\sigma^{2}}\right)^{i}\bigg|_{\sigma^{2}=0}\int dy\, f(y,x,\sigma^{2})=\int dy\,\left(\frac{\partial}{\partial\sigma^{2}}\right)^{i}\bigg|_{\sigma^{2}=0} f(y,x,\sigma^{2})
	\end{align}
	is well defined, and the power series in $\sigma^{2}$ of the integrand $f$ diverges for $|y|>C$ for some fixed $C>0$, then the power series in $\sigma^{2}$ of the integral $\hat{O}=\int dy\, f$ diverges as well. 
\end{lem} 
Though we were not able to rigorously prove this Lemma, the statement seems obvious to us.\footnote{For a general review of asymptotic series see \cite{Bender}.} Below, we show that both assumptions of this lemma hold for the real space correlation function and the 1D PDF.


\subsubsection*{Cosmological prerequisites: ZA and SPT for 1D correlators}\label{sec:}

To define the statistics of the displacement $  \psi(q) $ it is easiest to invert the Zel'dovich relation
\be\label{exact}
\delta(x)=\int d^{d}q \delta_{D}\left(  x-q-\psi(q)\right)\,,
\ee
and expand it to linear order, finding
\be\label{linearized}
\psi(q)=\int_{k}e^{ikq}\frac{ik}{k^{2}}\delta_{L}(k)\,,
\ee
where $  \delta_{L} $ is the linear order density and the integral is over $dk/(2\pi)$. Therefore
\be\label{notdelta}
\ex{\psi(q)\psi(0)}=\int_{k}e^{ikq}\frac{P_{L}(k)}{k^{2}}\,.
\ee
The variance of the 1D Zel'dovich displacement, which is exact before shell crossing, can then be computed
\begin{align}
\sigma^{2}(q)&=\langle\left[  \psi(q)-\psi(0)\right]^{2}\rangle=\int_{0}^{\infty}\frac{dk}{\pi}\frac{2P_{L}(k)}{k^{2}}\left[  1-\cos\left(  kq\right)\right]\label{defsig}\\
&=\ssi-\ssq(q)\,,
\end{align}
where we defined 
\begin{align}
\ssi&\equiv\int_{0}^{\infty}\frac{dk}{\pi}\frac{2P_{L}(k)}{k^{2}}>0\,,\\
\ssq(q)&\equiv\int_{0}^{\infty}\frac{dk}{\pi}\frac{2P_{L}(k)}{k^{2}}\cos\left(  kq\right)=\int_{-\infty}^{\infty}\frac{dk}{\pi}\frac{P_{L}(k)}{k^{2}}e^{iqk}\,,
\end{align}
and assumed that
\begin{align}
\ssq(q=0)=\ssi<\infty\,.
\end{align}
This property holds in a $  \Lambda $CDM universe where both the UV and IR part of the integral converge\footnote{For small $ k $, $ kP_{L}=k^{3}P_{3D}\sim k^{4} $, and so $ P_{L}\sim k^{3} $}. As we show in subsection \ref{LCDMpk} by marginally extending the proof of \cite{McQuinn:2015tva}, SPT converges to the ZA power spectrum. On the other hand, there exist (less realistic) cases in which $  \sigma(q)^{2} $ is unbounded, as for example in scaling universes, $  P_{L}(k)=A k^{n} $ for some $  n $. In subsection \ref{scalingpk} we discuss the finding of S. Foreman \cite{Simon} that SPT for the power spectrum can both converge or diverge depending on $  n $. Notice that $ P(k)=P(-k) $ and so $ \sigma^{2}(q) $ is real, as evident from \eqref{defsig}. Also, despite its name, $ \ssq $ does \textit{not} need to be positive, unlike $ \ssi $.

A standard result in Fourier analysis guarantees that if a function is square integrable, the Fourier transform vanishes at least as fast as $ q^{-1} $ for large $ q $. If the function is also continuous, then the Fourier transform vanishes as $ q^{-2} $. For a $  \Lambda $CDM-like universe we have that $  P(k)/k^{2} $ is a continuous, square integral function and therefore $ \ssq $ vanishes for large $ q $ as $ q^{-2} $, justifying the name of $ \ssi $.

\begin{figure}[h]
	\centering
	\includegraphics[width=1.2\columnwidth]{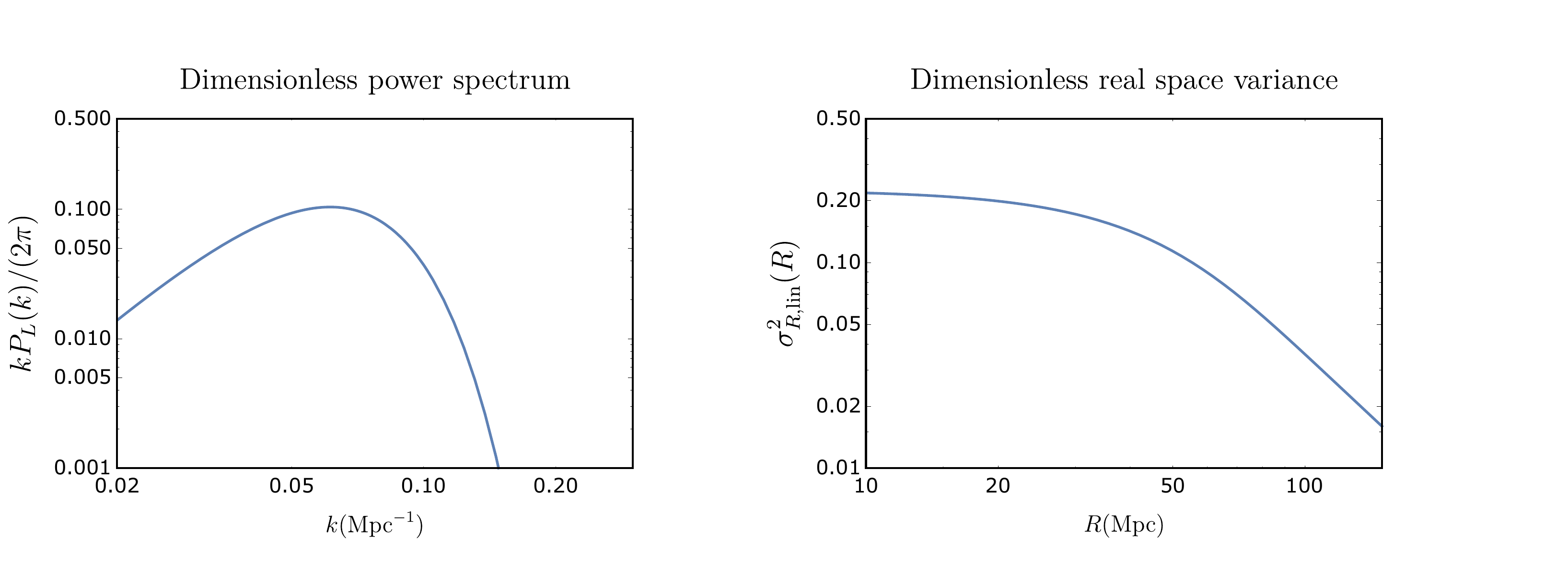}
	\caption{The plot shows the initial conditions that are used throughout this paper in two forms. In particular it shows that the dimensionless variance \eqref{dimvar} is significantly less than unity on all scales.\label{plotsvariance}}
\end{figure}

Throughout the paper, we support our analytical results with some plots of the observables in question. For simplicity, and in order to be least sensitive to multistreaming, we use the following linear power spectrum as initial condition in all our plots 
\begin{align}\label{exppk}
P_{L}(k)=\frac{4 \times 10^{4}}{\pi}k^{2}e^{-\frac{k^{2}}{0.05^{2}}}.
\end{align}
Its corresponding dimensionless variance for the average density in cells of size $R$ is given by
\begin{align}\label{dimvar}
\sigma_{R,\text{lin}}^{2}(R)=\int_{0}^{\infty}\frac{1}{\pi}\left(\frac{2}{kR}\right)^{2}\sin^{2}\left(\frac{kR}{2}\right)P_{L}(k)=\frac{2\times 10^{4}}{\pi^{3/2}}\frac{1-e^{-\frac{R^{2}}{1600}}}{R^{2}}=\frac{\sigma^{2}(R)}{R^{2}}.
\end{align}
To get an idea of the size of the perturbation parameter in this work, we plot $\sigma_{R,\text{lin}}^{2}$ and the dimensionless power spectrum, $kP(k)/2\pi$ in 1D, in figure \ref{plotsvariance}. 


\subsection{Convergence of SPT for the power spectrum for $  \Lambda $CDM-like universes}\label{LCDMpk}

Let us start our discussion considering a $  \Lambda $CDM-like 1D power spectrum as discussed above. Recall that the ZA expression for the power spectrum is \cite{Fisher:1995ec,McQuinn:2015tva}
\begin{align}\label{PZA}
P_{ZA}(k)=\int_{-\infty}^{\infty}dq e^{-ikq}\left[e^{-k^{2}\sigma^{2}(q)/2}-1\right].
\end{align}
Using the decomposition $\sigma^{2}(q)=\sigma_{\infty}^{2}-\sigma_{q}^{2}(q)$, this can be rewritten as 
\begin{align}\label{PStrick1}
P_{ZA}(k)=\int_{-\infty}^{\infty}dq e^{-ikq}\left[e^{-k^{2}\sigma_{\infty}^{2}/2}\left(e^{k^{2}\sigma_{q}^{2}(q)/2}-1\right)+\left(e^{-k^{2}\sigma_{\infty}^{2}/2}-1\right)\right].
\end{align}
The Fourier transform of the last term in brackets should be interpreted as a distribution, in which case we get
\begin{align}\label{PStrick2}
\int_{-\infty}^{\infty}dq e^{-ikq}\left(e^{-k^{2}\sigma_{\infty}^{2}/2}-1\right)=\left(e^{-k^{2}\sigma_{\infty}^{2}/2}-1\right)\delta_{D}(k)=0.
\end{align}
Thus we are left with 
\begin{align}\label{powerspectrum}
P_{ZA}(k)=\int_{-\infty}^{\infty}dq e^{-ikq}\left[e^{-k^{2}\sigma_{\infty}^{2}/2}\left(e^{k^{2}\sigma_{q}^{2}(q)/2}-1\right)\right].
\end{align}
For given $q$, the remaining term in brackets can be written as its Taylor series in $  \sigma^{2} $, where $\sigma_{\infty}$ and $\sigma_{q}$ count at the same order: 
\begin{align}
e^{-k^{2}\sigma_{\infty}^{2}/2}\left(e^{k^{2}\sigma_{q}^{2}(q)/2}-1\right)=\sum_{n=1}^{\infty}\left[\sum_{j=1}^{n}\frac{(k^{2}\sigma_{q}^{2}/2)^{j}}{j!}\frac{(-k^{2}\sigma_{\infty}^{2}/2)^{n-j}}{(n-j)!}\right].
\end{align}
In order to use Fubini-Tonelli, observe that 
\begin{align}
&\sum_{n=1}^{\infty}\left|\sum_{j=1}^{n}\frac{(k^{2}\sigma_{q}^{2}/2)^{j}}{j!}\frac{(-k^{2}\sigma_{\infty}^{2}/2)^{n-j}}{(n-j)!}\right| \leq \sum_{n=1}^{\infty}\sum_{j=1}^{n}\frac{|k^{2}\sigma_{q}^{2}/2|^{j}}{j!}\frac{|k^{2}\sigma_{\infty}^{2}/2|^{n-j}}{(n-j)!}\nonumber \\
&=e^{k^{2}\sigma_{\infty}^{2}/2}\left(e^{k^{2}|\sigma_{q}^{2}(q)|/2}-1\right),
\end{align}
where we have used that $\sigma_{\infty}^{2}>0$, but $\sigma_{q}^{2}$ is not necessarily positive. Thus,
\be
\int_{-\infty}^{\infty}dq \sum_{n=1}^{\infty}\left|\sum_{j=1}^{n}e^{-ikq}\frac{(k^{2}\sigma_{q}^{2}/2)^{j}}{j!}\frac{(-k^{2}\sigma_{\infty}^{2}/2)^{n-j}}{(n-j)!}\right|&\leq& \int_{-\infty}^{\infty}dq e^{k^{2}\sigma_{\infty}^{2}/2}\left(e^{k^{2}|\sigma_{q}^{2}(q)|/2}-1\right) \nonumber \\
&=&e^{k^{2}\sigma_{\infty}^{2}/2}\int_{0}^{\infty}dq \left(e^{k^{2}|\sigma_{q}^{2}(q)|/2}-1\right)\nonumber\\
&<&\infty\,,\nonumber
\ee
where in the last step we used that $\sigma_{q}^{2}$ goes to zero at least as $q^{-2}$ as $q \to \infty$. We conclude that Fubini-Tonelli can indeed be applied. Thus we find that SPT converges to the ZA expression,
\be\label{PSPT}
P_{ZA}(k)&=&\int_{-\infty}^{\infty}dq e^{-ikq} \sum_{n=1}^{\infty}\left[\sum_{j=1}^{n}\frac{(k^{2}\sigma_{q}^{2}/2)^{j}}{j!}\frac{(-k^{2}\sigma_{\infty}^{2}/2)^{n-j}}{(n-j)!}\right]\nonumber\\
&=&\sum_{n=1}^{\infty}\int_{-\infty}^{\infty}dq e^{-ikq}\left[\sum_{j=1}^{n}\frac{(k^{2}\sigma_{q}^{2}/2)^{j}}{j!}\frac{(-k^{2}\sigma_{\infty}^{2}/2)^{n-j}}{(n-j)!}\right]\nonumber \\
&=&\sum_{n=1}^{\infty}P_{n}(k)=P_{SPT}(k)\,.
\ee 

\begin{figure}[h]
	\centering
	\includegraphics[width=\columnwidth]{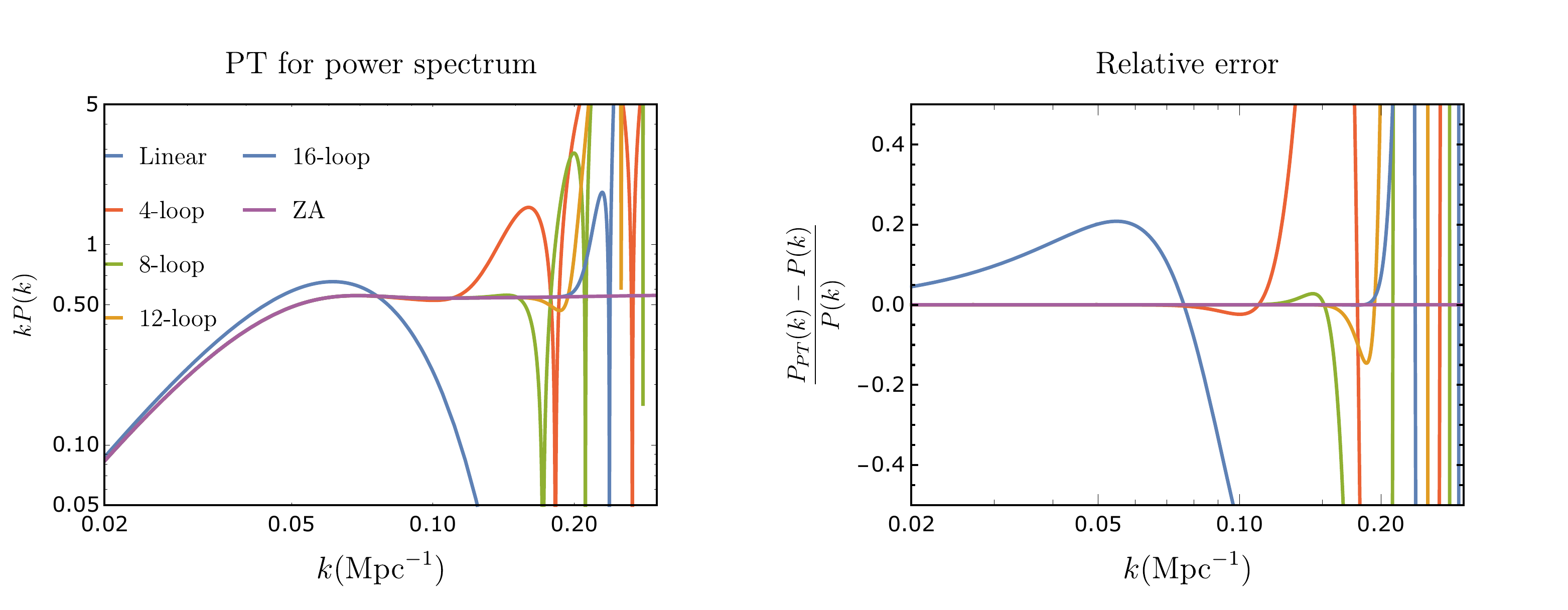}
	\caption{The plot shows the ZA power spectrum (purple continuous line) and some of its SPT approximations (full plot on left and relative error on right). For given $k$, one can reach arbitrary precision by going to high enough orders in perturbation theory, which is the hallmark of a convergent series. We use \eqref{exppk} as initial condition.\label{ptmcw}}
\end{figure}

To confirm this analytic result we can plot the power spectrum for initial condition \eqref{exppk}. The result is shown in Figure \ref{ptmcw}.


\subsection{Divergence of dimensionally regulated SPT for the power spectrum for scaling universes}\label{scalingpk}

Let us consider now scaling universes \cite{Peebles:1980}, $  P_{L}(k)=Ak^{n} $, for some $  n>-1 $. These avoid the conclusion of the previous subsection where we assumed $  \sigma^{2}<\infty $. The easiest case is actually $  n=0 $, so we discuss it first and then move to arbitrary $  n $. For $  n=0 $, the variance of the displacement can be computed analytically from \eqref{defsig} to be $  \sigma^{2}(q)=A|q| $. The ZA power spectrum follows from \eqref{PZA}:
\be
P_{ZA}(k)= \frac{2A}{4+ k^{2}A^{2}}\,.
\ee
SPT is an expansion in $  A $ and therefore has a finite radius of convergence $  |kA|=1 $. For some given $  A $, SPT converges on large scales but diverges on short scales. Moving on to arbitrary $  n $, we are forced to regulate the infinities. We quote the result of S. Foreman \cite{Simon}, who used dimensional regularization to obtain
\be
P_{L\text{-loop}}(k)=\frac{\pi^{L/2}}{(L+1)!}\frac{\Gamma(\frac12(n-1))^{L+1}\Gamma(\frac12(L-Ln-n+2))}{\Gamma(1-\frac n2)^{L+1}\Gamma(\frac12(-L+Ln+n-1))}\,\left(  \frac{k}{k_{NL}}\right)^{L(1+n)}P_{L}(k)\,,
\ee
where $  L $ is the number of loops and $  A=2\pi k_{NL}^{-1-n} $. Note that this result refers to the cut-off independent contribution, a.k.a. the ``finite part'' of the loop correction. To study the convergence of this series, let us focus on the Taylor coefficients $  a_{L} $ in the loop expansion
\be
P_{L\text{-loop}}(k)=a_{L}\left(  \frac{k}{k_{NL}}\right)^{L(1+n)}P_{L}(k)\,.
\ee
For large $  L $ one finds
\be
a_{L}\simeq \left(  \frac{c_{n}}{L^{n}}\right)^{L}\,
\ee
where $  c_{n} $ is some $  n $-dependent constant. We hence conclude that for $  -1<n<0 $ the series diverges for any $  k $. In this case the $  a_{L} $'s have indeed the typical growth encountered in divergent asymptotic expansions, which stop approaching the exact result at some $  k $-dependent order $  L_{\text{opt}} $. On the other hand, the  series converges for $  n>0 $ for any $  k $, since $  a_{L} $ decreases rapidly with $  L $. The boundary case $  n=0 $ was discussed above and has somewhat hybrid behavior converging only for some range of scales. 

Note that the dimensionally regulated results above show that the divergence of PT we discuss in this paper is indeed unrelated to the loop-divergences that are renormalized in the EFT of LSS \cite{Pajer:2013jj}. We further discuss this point in \ref{EFT}.


\subsection{Non-convergence of SPT for real space correlation function} \label{nonccf}

The two-point correlation function is obtained by Fourier transforming the power spectrum \cite{Carlson:2012bu},
\begin{align}
\xi(r)=\int_{-\infty}^{\infty}\frac{dk}{2\pi}e^{ikr}P(k).
\end{align}
Applying this to \eqref{powerspectrum}, we just need to evaluate Gaussian integrals, and find 
\begin{align}\label{corrfnc}
1+\xi(r)=\int_{-\infty}^{\infty}dq \frac{1}{\sqrt{2\pi\sigma^{2}(q)}}e^{-\frac{(r-q)^{2}}{2\sigma^{2}(q)}}.
\end{align}
Obviously, even if we forget about the square root factors, the Taylor expansion of the integrand around $  \sigma^{2}(q)=0 $ does not converge to the right result. Hence, a convergence proof such as the one above does not exist. In fact, in the following, we argue that SPT does not converge to ZA. To that end, we show that this expression satisfies the conditions of Lemma \ref{assumption}. More precisely, we show that upon a change of variables, the integrand contains an essential singularity in the domain of integration. By defining $  q=\sigma_{\infty}x+r $, we can write \eqref{corrfnc} as\footnote{One might worry about the validity of this change of coordinates when $\sigma_{\infty}=0$ (which is in fact the point we wish to perturb around). However, it suffices to show that \eqref{corrfnc} and \eqref{corrfnc2} are identical as a function of $r$ and $\sigma_{\infty}$. This is obvious for finite $\sigma_{\infty}$, so we just need to investigate both functions in the limit $\sigma_{\infty}\to 0$. In order to see this, observe that in this limit, the integrand in \eqref{corrfnc} simply becomes a $\delta_{D}(r-q)$, and the integrand in \eqref{corrfnc2} just a Gaussian with constant variance $\sigma^{2}(r)/\sigma_{\infty}^{2}$ (this ratio does not scale with $\sigma_{\infty}$). Fortunately, integrating over both just gives unity, such that, as expected, in this limit the correlation function is just zero. Thus we have shown that these expressions are indeed equivalent. We choose to analyze the latter.}
\begin{align}\label{corrfnc2}
1+\xi(r)=\int\frac{dx}{\sqrt{2\pi}}\frac{1}{\sqrt{\sigma^{2}(\sigma_{\infty}x+r)/\sigma_{\infty}^{2}}}e^{-\frac{x^{2}}{2}\frac{1}{\sigma^{2}(\sigma_{\infty}x+r)/\sigma_{\infty}^{2}}}.
\end{align}
Note that since $\sigma^{2}(q)$ and $\sigma_{\infty}^{2}$ are of the same order in our perturbation parameter, we only need to expand with respect to the $\sigma_{\infty}^{2}$ in the argument. More explicitly, let's write
\begin{align}
\sigma^{2}(q)/\sigma_{\infty}^{2}=1-\sigma_{q}^{2}/\sigma_{\infty}^{2}=1-f(q), 
\end{align}
for which typically $|f(q)|\leq 1$. The perturbative series is then obtained in the standard way:
\begin{align}\label{xipt}
\xi_{n}(r)=\frac{1}{n!}\left(\frac{\partial}{\partial \sigma_{\infty}}\right)^{n}\int\frac{dx}{\sqrt{2\pi}}\frac{1}{\sqrt{1-f(\sigma_{\infty}x+r)}}e^{-\frac{x^{2}}{2}\frac{1}{1-f(\sigma_{\infty}x+r)}}\Bigg|_{\sigma_{\infty}=0}.
\end{align}
One can check that this gives the same series as before \textit{upon interchanging the order of the integral and derivatives}. 

In order to use Lemma \ref{assumption}, we analyze the integrand. More precisely, we show below that the point $\sigma_{\infty}=-r/x$ is an essential singularity in the complex $\sigma$-plane. The series therefore diverges for $\sigma_{\infty}>|r/x|$. Conversely, this means that for given $\sigma_{\infty}$, the series diverges for $x>|r/\sigma_{\infty}|$. Moreover, the integral of derivatives of the integrand evaluated at $\sigma=0$ are well defined. Hence we conclude that SPT diverges for the correlation function, given the non-convergence proof below.  

\begin{figure}[h!]
	\centering
	\includegraphics[width=\columnwidth]{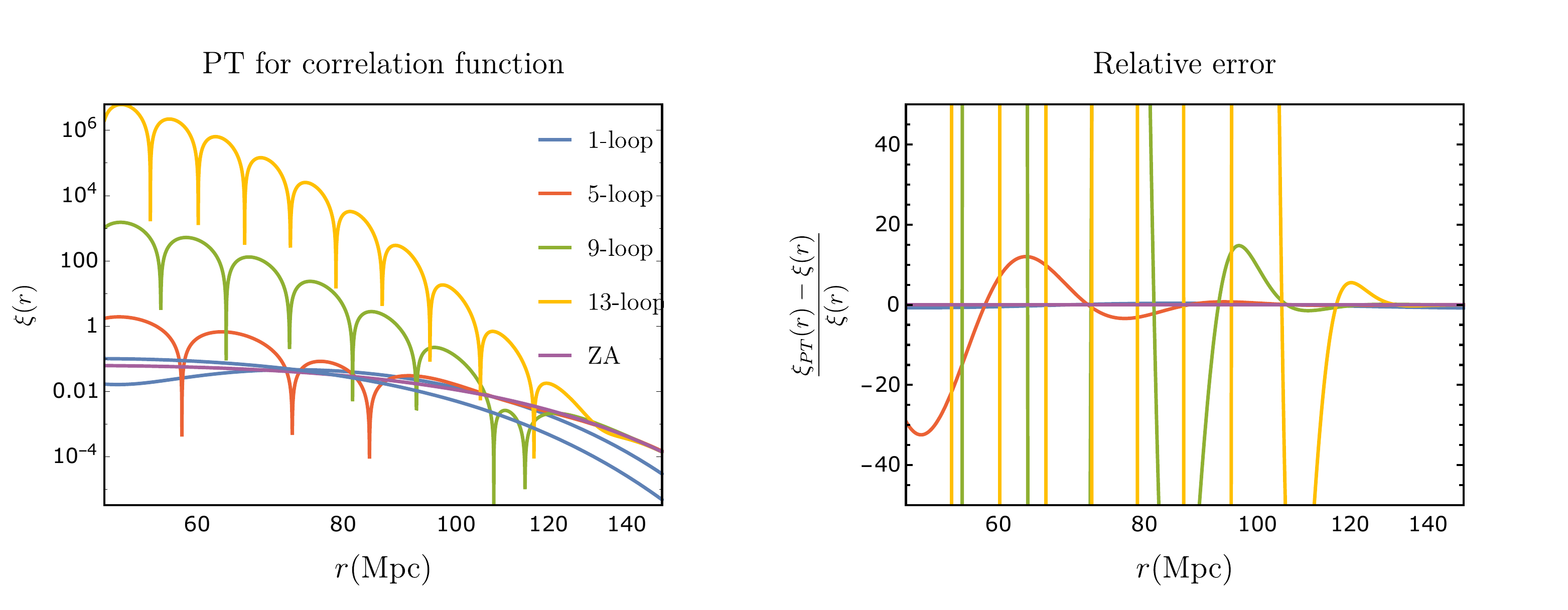}
	\caption{The plot shows the ZA correlation function and some of its SPT approximations. At any given $r$, perturbation theory stops improving and starts diverging from the exact result at high enough orders. At larger scales the divergence sets in at a higher loop order. This provides numerical evidence for the non-convergence of the SPT series for real space correlators, for arbitrarily large $r$. We use \eqref{exppk} as initial power spectrum.\label{figcf}}
\end{figure}

\begin{proof}
Let us define
\begin{align}
I(x,\sigma,r)=\frac{1}{\sqrt{1-f(\sigma_{\infty}x+r)}}e^{-\frac{x^{2}}{2}\frac{1}{1-f(\sigma_{\infty}x+r)}},
\end{align} 
we ask whether
\begin{align}
\sum_{n=0}^{\infty}\frac{1}{n!}\left(\frac{\partial }{\partial \lambda}\right)^{n} I(x,\lambda,r)\bigg|_{\lambda=0}\sigma^{n}\overset{?}{=}I(x,\sigma,r),
\end{align}
for every $x$, and given $r$. Let us now define a slight modification of this function
\begin{align}
\tilde{I}(x,\lambda,r)\equiv I(x,\sigma/x,r)=\frac{1}{\sqrt{1-f(\lambda +r)}}e^{-\frac{x^{2}}{2}\frac{1}{1-f(\lambda+r)}}.
\end{align} 
This has the property that 
\begin{align}
\sum_{n=0}^{\infty}\frac{1}{n!}\left(\frac{\partial }{\partial \lambda}\right)^{n} \tilde{I}(x,\lambda,r)\bigg|_{\lambda=0}(\sigma x)^{n}=\sum_{n=0}^{\infty}\frac{1}{n!}\left(\frac{\partial }{\partial \lambda}\right)^{n} I(x,\lambda,r)\bigg|_{\lambda=0}\sigma^{n}.
\end{align}
This is useful for the following reason. Since $f(0)=1$ by definition, and assuming $f$ to be smooth and symmetric around zero, this means 
\begin{align}
\left(\frac{\partial }{\partial \lambda}\right)^{n} \tilde{I}(x,\lambda,r)\bigg|_{\lambda=-r}=0,
\end{align}  
so this function is certainly non-analytic in $\lambda$ at $\lambda=-r$. The radius of convergence for $\tilde{I}$ around $0$ is at most $r$, since this point constitutes an essential singularity in the complex plane. In fact, it is very reminiscent of the expansion of $e^{-1/(1+x)^{2}}$ around zero, which has radius of convergence $1$. Thus we conclude that for $|x|>r/\sigma$,
\begin{align}
\sum_{n=0}^{\infty}\frac{1}{n!}\left(\frac{\partial }{\partial \lambda}\right)^{n} \tilde{I}(x,\lambda,r)\bigg|_{\lambda=0}(\sigma x)^{n}\neq I(x,\sigma,r) \quad \text{for }|x|>r/\sigma.
\end{align}
But we can rewrite the left hand side back in its original form (this is a strict equality that follows from its definition), and conclude
\begin{align}
&\sum_{n=0}^{\infty}\frac{1}{n!}\left(\frac{\partial }{\partial \lambda}\right)^{n}x^{n} \tilde{I}(x,\lambda,r)\bigg|_{\lambda=0}\sigma^{n}=\nonumber \\
&=\sum_{n=0}^{\infty}\frac{1}{n!}\left(\frac{\partial }{\partial \lambda}\right)^{n} I(x,\lambda,r)\bigg|_{\lambda=0}\sigma^{n}\neq I(x,\sigma,r) \quad \text{for }|x|>r/\sigma.
\end{align}
\end{proof}
Thus we conclude that perturbation theory for the correlation function diverges. Assuming convergence for $|x|<r/\sigma$, we can estimate the error as the contribution to the integral of the `tails': the collection $x$, for which $|x|>r/\sigma$. Since $f$ goes to zero for large arguments, these tail contributions are roughly exponentially suppressed by $e^{-(r/\sigma)^{2}}$, because of the exponent in the integrand. To be more precise, upon expanding, every term will be evaluated at $\sigma_{\infty}=0$. This means the exponential suppression is $e^{-\frac{x^{2}}{2\sigma^{2}(r)/\sigma_{\infty}^{2}}}$. Plugging in the relevant value of $x$, we find that the non-perturbative error is indeed exponentially suppressed in the dimensionless variance 
\begin{align}
\text{NP-error} \sim e^{-\left(\frac{r}{\sigma(r)}\right)^{2}}. 
\end{align}  
We can verify numerically the non-convergence using the initial power spectrum \eqref{exppk}. Figure \ref{figcf} shows the comparison for the correlation function of the ZA result and the SPT approximations. For example, around $  r=90 \text{ Mpc} $, we see that the 5 loop result is much closer to ZA than the 9 loop one. At those scales the series stops diverging somewhere between 5 and 9 loops. The divergence of perturbation theory at larger scales shows up at higher order in PT because the variance is smaller. For example, at $  r\sim 110 \text{ Mpc}$, PT is getting closer to the right answer up to 9 loops but then start diverging somewhere between 9 and 13 loops.

Finally, one could ask why a convergence proof starting from the convergent series for the power spectrum does not work. The reason is that the $P_{n}$ are not all positive. We comment on this further in Appendix \ref{app:RPT}.


\subsection{Convergence of SPT for the Bispectrum and NG initial conditions}

The bispectrum in 1D can be intuitively expressed as
\begin{align}\label{bispectrum}
B(k_{1},k_{2})=V^{-1}\langle|\delta(k_{1})\delta(k_{2})\delta(-k_{1}-k_{2})|\rangle,
\end{align}
where $V$ is the 1D spatial volume. In Appendix \ref{ssec:tt}, we derive this expression more formally from symmetries. Plugging in the ZA expressions for the density, we obtain
\begin{align}
B(k_{1},k_{2})=V^{-1}\langle\int & dq_{123}e^{-i k_{1}q_{1}}e^{-i k_{2}q_{2}}e^{i (k_{1}+k_{2})q_{3}}\nonumber\\
&\times\left(e^{-ik_{1}\Psi_{1}}-1\right)\left(e^{-ik_{2}\Psi_{2}}-1\right)\left(e^{i(k_{1}+k_{2})\Psi_{3}}-1\right)\rangle,
\end{align}
where $\Psi_{i}\equiv\Psi(q_{i})$. In Appendix \ref{app:bispectrum}, we work out this expression and show that for Gaussian initial conditions, perturbation theory converges for the bispectrum. 

For NG initial conditions, we use the cumulant expansion theorem as in equation (2.27) of \cite{McQuinn:2015tva}, to find an expression for the power spectrum of the form
\begin{align}
P^{NG}(k)=\int_{-\infty}^{\infty}dq e^{-ikq}\left[e^{-k^{2}\sigma^{2}(q)/2+ik^{3}\sigma_ {3}(q)/3!-k^{4}\sigma_{4}(q)/4!+\ldots}-1\right]\,,
\end{align}
where $  \sigma_{3,4,\dots} $ characterize the type of primordial non-Gaussianity. Then, depending on the behavior of the $\sigma_{n}$, one can show the convergence of PT in a similar fashion. Once again, we refer the interested reader to Appendix \ref{app:ng} for details.


\section{The 1D count-in-cell PDF: non-convergence of PT for cumulants}\label{sec:4}

In this section we discuss cumulants obtained from the probability distribution function (PDF) for the density averaged in cells of a certain fixed radius $R$. This observable recently attained some renewed interest in the 3D context, see e.g. \cite{Bernardeau:2013dua,Uhlemann:2015npz}, but has been object of study for quite some time \cite{Bernardeau:1992zw,Bernardeau:1993qu,Bernardeau:1994aq,Bernardeau:1994zd,Bernardeau:2000et,Kofman:1993mx,Protogeros:1996ex,Fosalba:1997tn,Ohta:2004mx,Gaztanaga:1998mw,Lam:2007qw}. Here we show that in 1D, it is possible to use the exact solution for planar collapse, see Appendix \ref{planar}, to compute this PDF exactly up to shell crossing events (which are limited for our choice of initial conditions \eqref{powerspectrum}). As we will see, the application of the construction of the PDF in 3D, which relies on spherical collapse, is easily applied to 1D. Most approaches \cite{Bernardeau:2000et,Bernardeau:2013dua,Fosalba:1997tn,Lam:2007qw} agree on the exponential behavior of the PDF, but there is still some confusion about the prefactor \cite{Lam:2007qw}, which is related to normalization constraints on the PDF: it should be unitary and have vanishing first moment. Typically normalization is enforced by hand \cite{Kofman:1993mx,Protogeros:1996ex,Ohta:2004mx,Lam:2007qw}. Here we present a new derivation of the so called Lagrangian space PDF, and show how, at least in 1D, one can also obtain the final Eurlerian space PDF from first principles, which automatically satisfies the unitarity and mass conservation contraints.  
Moreover, we show how the tails of this distribution are beyond the reach of perturbation theory, implying a non-perturbative error in the computation of any cumulant. This was also observed in the 3D context in \cite{Valageas:2009km}. We formalize the argument for non-convergence and highlight the relation to the non-perturbative error for the correlation function obtained in \ref{nonccf}.   

\subsection{Conservation of probability} 

Let us start with the definition of the PDF for the density averaged in cells. The question the PDF should answer is the following:
\begin{itemize}
	\item If one picks a random point $x$ in space, what is the probability $P[\bar{\delta}_{R}]d\bar{\delta}_{R}$that the average density between $x$ and $x+R$, which we call $\delta_{R}(x)$, is in the range $[\bar{\delta}_{R},\bar{\delta}_{R}+d\bar{\delta}_{R}]$?
\end{itemize}
Assuming ergodicity, an equivalent but more useful way to phrase the question is
\begin{itemize}
	\item What fraction of the spatial volume $x\in V$ has the property that $\delta_{R}(x)$ is in the range $[\bar{\delta}_{R},\bar{\delta}_{R}+d\bar{\delta}_{R}]$?
\end{itemize}
One can try to compute the answer to this question in two steps. The first step is to find how the above property translates into a property in the initial conditions: if $\delta_{R}(x)$ is in the range $[\bar{\delta}_{R},\bar{\delta}_{R}+d\bar{\delta}_{R}]$, what does this mean for $\delta_{i,R_{i}}(q(x))$? Here $q(x)$ is the initial position of the fluid element that ends up at $x$ and $\delta_{i},R_{i}$ are to be found from the cell dynamics. Since we know the statistics of the initial conditions, we can then compute fraction of the initial volume with this property. The second step is to compute how the volume fraction $q\in V_{i}$ with this property changes as a function of time. Combining these steps allows us to find the PDF.

\subsection{Lagrangian space PDF}

For simplicity and clarity, we start by forgetting about the second step, which has the same main features in terms the qualitative conclusions we like to draw from the PDF. We call it the Lagrangian PDF, and it gives the following probability:
\begin{itemize}
	\item What fraction of the initial volume $q\in V_{i}$ has the property that at the final time $\delta_{R}(x(q))$ is in the range $[\bar{\delta}_{R},\bar{\delta}_{R}+d\bar{\delta}_{R}]$?
\end{itemize}
There are two ways to obtain this Lagrangian PDF. The first is more elaborate and insightful, and, to the best of our knowledge, it is novel. The second is more elegant mathematically and can already be found in the literature \cite{Fosalba:1997tn,Lam:2007qw,Bernardeau:2001qr}. We present them both here. 

\subsubsection*{Method 1}

Let us use the equations of motion to consider what the range $[\bar{\delta}_{R,f},\bar{\delta}_{R,f}+d\bar{\delta}_{R,f}]$ maps into in the initial conditions. The equations are the (radius independent) equation for the evolution of the density, supplemented with a mass conservation equation,
\begin{align}\label{masscons}
1+\delta_{f}&=\frac{1}{1-\delta_{L}},\nonumber \\
(1+\delta_{f})R_{f}&=R_{i} ,
\end{align}
where $\delta_{L}=\frac{a}{a_{i}}\delta_{i}$. This means that this range in $\delta$ at fixed $R_{f}$, maps into a line in $\{\delta,R_{i}\}$-space in the initial  conditions (in the linear approximation). We care about its slope $s$: 
\begin{align}
s=\frac{d\delta_{L}}{dR_{i}}=\frac{d\delta_{L}}{d\delta_{f}}\frac{d\delta_{f}}{dR_{i}}=\frac{1}{(1+\delta_{f})^{2}R_{f}}.
\end{align}
Thus the question becomes what the probability is that, given some point $q$ in the initial conditions, the function $\delta_{R_{i}}(q)\equiv \delta_{L}(R_{i})$ (it better be $q$-independent) as a function of $R_{i}$ crosses the infinitesimal line element $l$ between the points $\{R_{i},\delta_{L}\}$ and $\{R_{i}+d\delta_{L}/s,\delta_{L}+d\delta_{L}\}$ somewhere. Here $d\delta_{L}=d\delta_{f}/(1+\delta_{f})^{2}$. For this, we need the following (correlated) ingredients:
\begin{enumerate}
	\item What is the initial probability distribution for the density in cell of length $R_{i}$, $P_{R_{i}}(\delta_{L})$? 
	\item What is the initial probability distribution for the derivative $\mu\equiv \frac{d\delta_{L}(R_{i})}{dR_{i}}$, given $\delta_{L}(R_{i})$? 
\end{enumerate}
In other words, we need their joint PDF. Fortunately, we can assume Gaussian initial conditions, for which the joint PDF is merely a two-dimensional Gaussian, determined by the correlators $\langle\delta_{R_{i}}^{2} \rangle,\langle\delta_{R_{i}} \delta_{R_{i}}^{\prime}\rangle$, and $\langle\delta_{R_{i}}^{\prime 2} \rangle$. Here we used $\delta_{R_{i}}^{\prime}$ instead of $\mu$. Formally, we now have to integrate over all combinations of $\delta_{R_{i}}$ and $\delta_{R_{i}}^{\prime}$ such that the line $\delta_{R_{i}}+\lambda\delta_{R_{i}}^{\prime}$, where $\lambda \in {\rm I\!R}$, crosses $l$. And multiply its probabilities. The computation and corresponding approximations we leave to the appendix. The result for small enough variances however agrees with the much simpler expression we find next. The result is schematically given by 
\begin{align}
P_{R_{f}}(\delta_{f})= \text{Prefactor}(\delta_{f})P_{G}\left(\delta_{L}(\delta_{f}),\sigma_{R_{i}(\delta_{f},R_{f})}\right),
\end{align} 
where the prefactor is some non-exponential function of $\delta_{f}$, and $P_{G}$ is the Gaussian probability density. The main qualitative features are determined by the Gaussian. 

\subsubsection*{Method 2}

The key observation for this method is that the fundamental variable for the Lagrangian PDF, $y\equiv\delta_{L}/\sigma_{R_{i}}$, is in fact Gaussian distributed \cite{Fosalba:1997tn,Lam:2007qw}. Thus,
\begin{align}
P(y)dy=P_{G}(y,\sigma=1)dy,
\end{align} 
meaning
\begin{align}
P^{L}_{R_{f}}(\delta_{f})d\delta_{f}=P_{G}\left(\delta_{L}/\sigma_{R_{i}}\right)\times\left(\frac{d\delta_{L}}{d\delta_{f}}\frac{1}{\sigma_{R_{i}}}-\frac{\delta_{L}}{\sigma_{R_{i}}^{2}}\frac{d\sigma_{R_{i}}}{d\delta_{f}}\right)d\delta_{f},
\end{align}
where the expression on the right should be interpreted as a function of $R_{f}$ and $\delta_{f}$, using the mapping. This gives the same prefactor as the first method for small variances, corrections to which are negligible in our case. 

\subsection{Eulerian space PDF}

Even though the transition from Lagrangian to Eulerian space does not qualitatively change the PDF too much, and is therefore not too relevant for the purpose of this paper, we discuss it here for two reasons. First, we want to verify the statement that indeed the transition does not change the qualitative behavior too much. Second, as advertised above, there has been some discussion in the literature about the prefactor for the count-in-cell PDF in 3D, which is very similar in spirit to our 1D PDF. Our derivation of the Eulerian prefactor might inspire a new approach in that context as well. 

Method 1 to derive the Lagrangian PDF gives us a way to calculate the volume fraction satisfying the given property as an integral over the probability of all initial conditions that satisfy that property. The Eulerian density is obtained from the Lagrangian one by multiplying it by the ratio of the final volume to the initial volume that satisfies this property. Remember that the initial volume we are talking about is the collection of $q\in V_{in}$ satisfying a certain property. Assuming some continuity conditions, this is indeed a volume. Namely, if $\bar{q}$ satisfies the property, then at least an infinitesimal region around $\bar{q}$ does also. The evolution of this volume element is then determined by the local density at $\bar{q}$ in the standard Zel'dovich manner:
\begin{align}
r=\frac{dV_{f}\left(\bar{x}(\bar{q})\right)}{dV_{i}\left(\bar{q}\right)}=1-\delta_{L}(\bar{q}).
\end{align}
The probability density can thus be thought of as a weighted (by the joint probability) integral over all initial conditions that satisfy the Lagrangian condition, multiplied by this ratio $r$. Schematically, we can thus write
\begin{align}\label{PDFrough}
P_{R_{f}}(\delta_{f})d\delta_{f}=\int_{I|_{\text{Lagrangian property}}}P_{MVG}\left[\delta_{L}(q),\delta_{R_{i}}(q),\delta_{R_{i}}^{\prime}(q)\right](1-\delta_{L}(q)),
\end{align} 
where MVG stand for multi-variate Gaussian (as the initial conditions are Gaussian), $I$ stands for initial conditions, which in this case this means integration over the random variables $\{\delta_{L}(q),\delta_{R_{i}}(q),\delta_{R_{i}}^{\prime}(q)\}$. The sub-text indicates the restriction on them. Even though we wrote the $q$ dependence everywhere, statistical homogeneity guarantees the answer will not depend on it. Note that the multi-variate Gaussian depends on all non-vanishing cross correlations among the arguments as well. The detailed expression for this PDF can be found in appendix \ref{PDF}. In Figure \ref{variouspdfs} we plot the PDF for a couple of radii, and check its normalization conditions, which are all correct at the subpercent level.

\begin{figure}[h]
	\centering
	\includegraphics[width=\columnwidth]{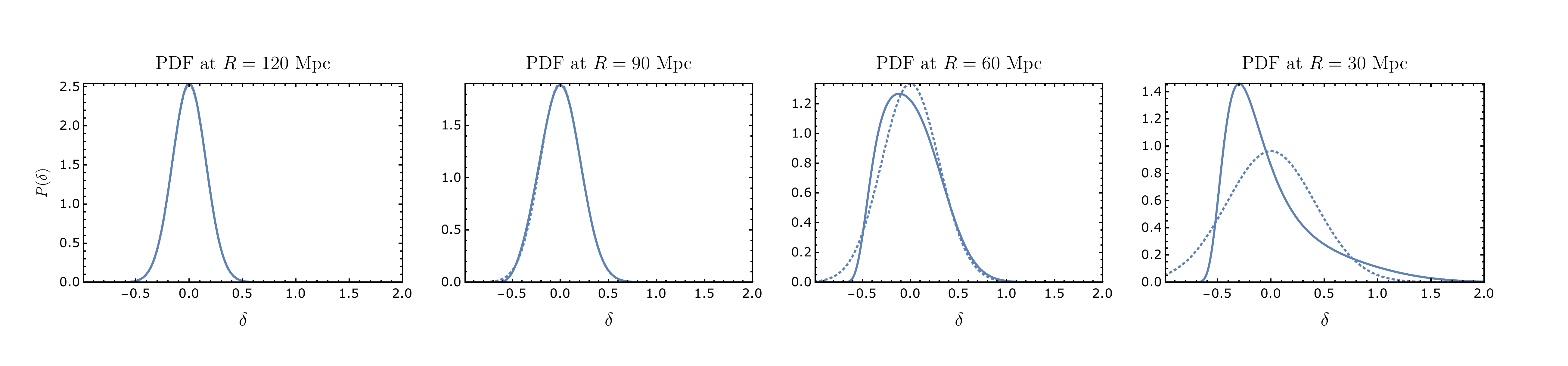}
	\caption{The plot shows the PDF \eqref{finalPDF} as a function of the final density evaluated at four different scales, respectively $R=120,90,60,30 \text{ Mpc}$, including the linear, Gaussian PDF in dotted blue for reference. Its norm and mean evaluate to $(1.006,3\times 10^{-5}),(1.005,2\times 10^{-4}),(1.004,7\times 10^{-4}),(1.004,2\times 10^{-3})$, respectively. We use \eqref{exppk} as initial condition.  \label{variouspdfs}}
\end{figure}

\subsection{Perturbation theory and convergence}

\begin{figure}[h]
	\centering
	\includegraphics[scale=0.8]{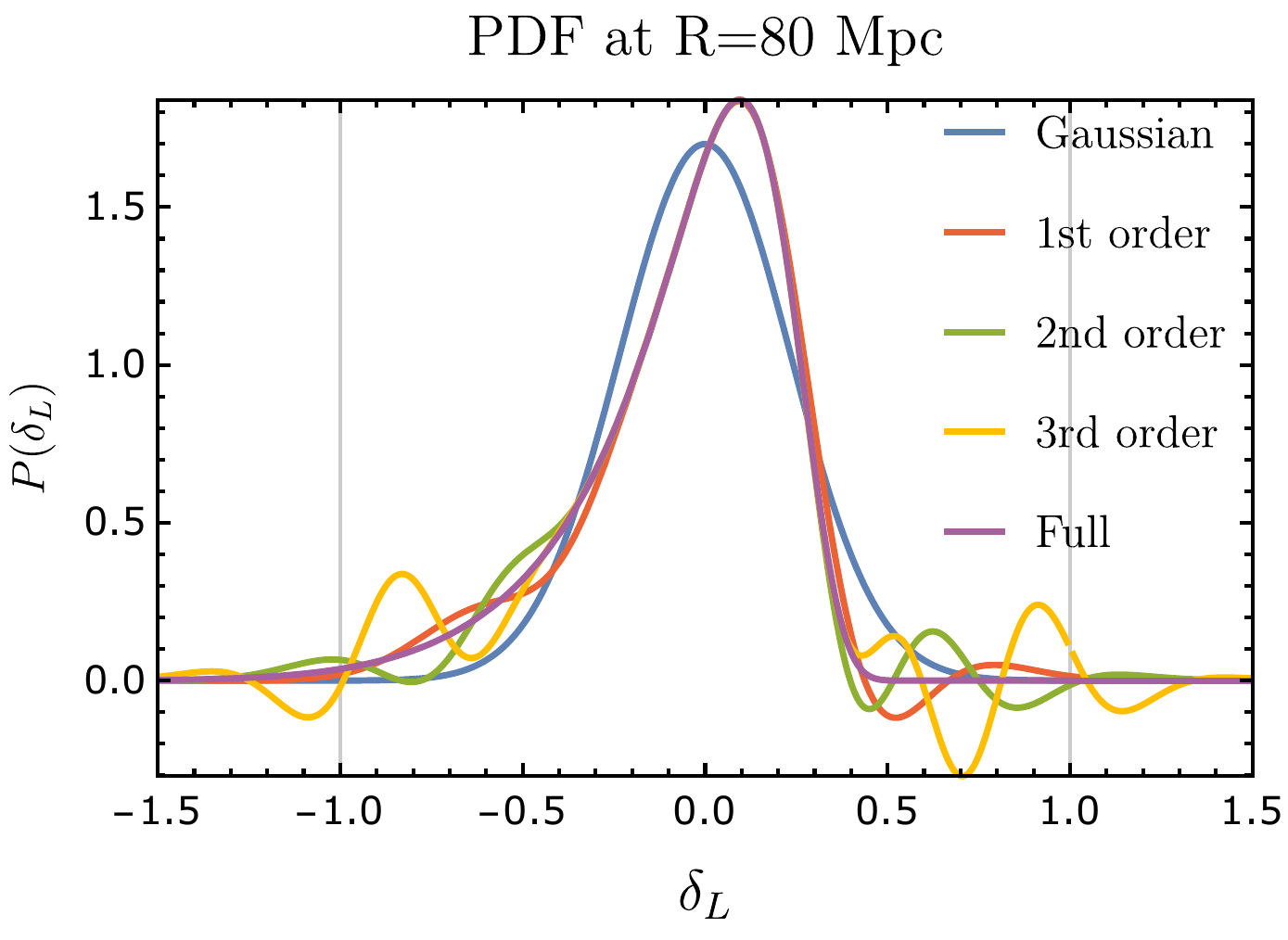}
	\caption{The plot shows the PDF for the average over cells of radius $  R=80 $ Mpc (purple continuous line) as well as some of its perturbative approximations. Perturbation theory captures the PDF pretty well around the peak and one can see that including higher order terms extends the region where PT agrees with the full result. The convergence is not expected to improve beyond radius of convergence, which is indicated by the grey vertical lines. We use \eqref{exppk} as initial power spectrum. \label{pdfplots}}
\end{figure}

Perturbation theory for the PDF proceeds similar to perturbation theory for the correlation function. As in the previous sections, perturbation theory here means a series in the amplitude of the primordial, linear power spectrum. As indicated in appendix \ref{PDF}, we can write the PDF solely as a function of $\delta_{L}$,
\begin{align}
P_{R_{f}}(\delta_{f})d\delta_{f}= \text{Prefactor}(\delta_{L})P_{G}\left(\delta_{L},\sigma_{R_{i}(\delta_{L},R_{f})}\right)d\delta_{L},
\end{align}
where $\sigma_{R_{i}(\delta_{L},R_{f})}^{2}=C_{22}(\delta_{L})$. This PDF is, a priori, not well defined beyond $\delta_{L}=1$, but the only sensible way to extend it is to set it to zero for larger values of the density, such that effectively we only integrate up to $\delta_{L}=1$, which corresponds to $\delta=\infty$. The perturbative expansion is most easily obtained by introducing an artificial small parameter $\lambda$ that multiplies the variance, and subsequently changing variable to $\tilde{\delta}=\delta_{L}/\lambda$. Due to the nature of the Gaussian distribution, $\lambda$ drops out in several places, and the PDF becomes
\begin{align}\label{PDFforPT}
P_{R_{f}}(\delta_{f})d\delta_{f}= \text{Prefactor}(\lambda\tilde{\delta},R_{f})P_{G}\left(\tilde{\delta},C_{22}(\lambda\tilde{\delta},R_{f})\right)d\tilde{\delta}.
\end{align}
The perturbative expansion is then obtained by expanding this function around $\lambda=0$, before evaluating it at $\lambda=1$. The convergence properties are most easily understood by realizing that both the prefactor and $C_{22}(\lambda\tilde{\delta},R_{f})$ in \eqref{PDFforPT} are actually functions of $(1-\lambda \tilde{\delta})$. Let us forget about the prefactor for simplicity. It is instructive to write out the Gaussian part more explicitly. Let us denote $\sigma_{R_{i}}^{2}=h(R_{i})$, which, through \eqref{masscons}, can be written as $h(R_{f}/(1-\delta_{L}))$. The Gaussian then becomes
\begin{align}
P_{G}\left(\tilde{\delta},C_{22}(\lambda\tilde{\delta},R_{f})\right)=\frac{1}{\sqrt{2\pi}}\frac{1}{\sqrt{h(R_{f}/(1-\lambda \tilde{\delta}))}}e^{-\frac{x^{2}}{2}\frac{1}{h(R_{f}/(1-\lambda \tilde{\delta}))}}.
\end{align}
Now note that for large $R_{i}$, $\sigma_{R_{i}}^{2}$ typically goes as $R_{i}^{-n}$ for $n>1$, implying $h(R_{f}/(1-\lambda \tilde{\delta}))\sim (1-\lambda \tilde{\delta})^{n}/R_{f}^{n}$. This means that, analogously to the expression for the real space correlation function, the perturbative expansion has an essential singularity at $\lambda = 1/\tilde{\delta}$. Conversely, setting $\lambda$ to unity, this means perturbation theory diverges beyond $|\tilde{\delta}|>1$. Since cumulants are obtained by integrating over the full PDF, perturbation theory for all cumulants is generically divergent (once again, no divergences appear in the limit $\lambda\to 0$, so we can intechange derivatives with respect to $\lambda$ and the integral). We plot the perturbative approximations to the PDF for $R_{f}=80\text{ Mpc}$ in Figure \ref{pdfplots}. This is similar in spirit to the asymptotic nature of Edgeworth expansions of PDFs \cite{Sellentin:2017aii}.

\newpage


\section{General properties of the non-perturbative error}\label{sec:5}

The previous sections contained concrete, explicit examples of the reach of perturbation theory for particular observables in particular settings. In contrast, we now speculate on the qualitative lessons we can learn from those examples. To highlight the generic nature of non-perturbative effects in real space, we review some well known facts about the halo model in appendix \ref{halo}, whose non-perturbative effects have been estimated in, e.g. \cite{Valageas:2013hxa,Afshordi:2006ch}. 

\subsection{Non-perturbative errors in real space}

\begin{figure}[h!]
	\centering
	\includegraphics[scale=0.8]{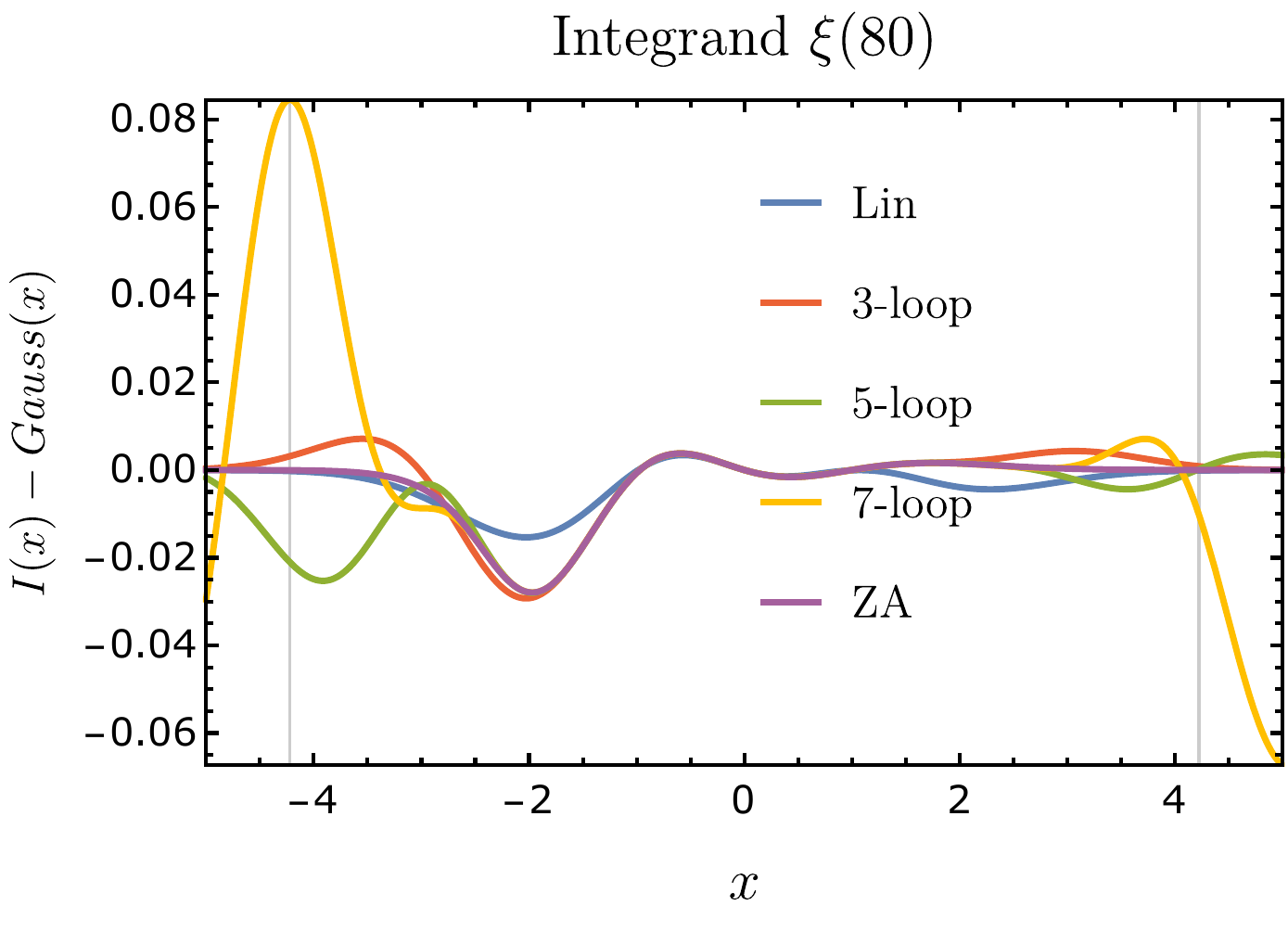}
	\caption{Here we plot the perturbative integrand for $\xi(80)$ minus the Gaussian whose integral evaluates to unity, and compare to the full result in red. Similar to the perturbative expression for the PDF, we see that perturbation theory performs well around the origin, but diverges wildly beyond the radius of convergence, $x=\pm r/\sigma_{\infty}$, which we plot as vetical grey lines. We use \eqref{exppk} as initial power spectrum.\label{nperrorplot2}}
\end{figure}

From the examples studied above, we can get some idea on a lower limit on the size of the non-perturbative error in real space. At least for the count-in-cell statistics and the correlations function, we found that observables $\mathcal{O}$ can be written qualitatively as an integral of the form
\begin{align}
\mathcal{O}(r)\sim \int\frac{dx}{\sqrt{2\pi}}\frac{1}{\sqrt{g_{r}(1-\lambda x)}}e^{-\frac{x^{2}}{2}\frac{1}{g_{r}(1-\lambda x)}},
\end{align}
for some function $g_{r}$, satisfying $g_{r}(0)=0$. As a reminder, $  \lambda $ is a fictitious coupling constant which we expand around $  \lambda=0 $, and which should be set to unity at the end of the calculation. We argued that the non-perturbative error comes from the $|x|\geq |1/\lambda|$ tails, since this point constitutes an essential singularity in the complex plane. This can be seen by approaching the point in the complex direction $\lambda x=1\pm i\epsilon$, where $\epsilon\to 0$. Thus, the error can be estimated as the tail contribution to each term in perturbation theory (PT):
\begin{align}\label{nperror}
\int_{\frac{1}{\lambda}}^{\infty}dx\, h(r,x)\frac{1}{\sqrt{2\pi g_{r}(1)}}e^{-\frac{x^{2}}{2g_{r(1)}}},
\end{align}  
for $h(r,x)$ some polynomial in $x$. At higher orders in PT the polynomial $  h $ becomes larger and larger, increasing the error on the tails. This can be seen explicitly in Figure \ref{nperrorplot2}, where we show the PT approximations of the integrand of the correlation function $\xi(r=80 \text{ Mpc})$, \eqref{xipt}. For clarity, we have subtracted the $0$-th order Gaussian, which is responsible for the $1$ in $1+\xi(r)$. We have indicated the values beyond which the integrand does not converge to the true answer anymore, given by $x=\pm r/\sigma_{\infty}$. 

\begin{figure}[h!]
	\centering
	\subfigure{\includegraphics[width=0.8\columnwidth]{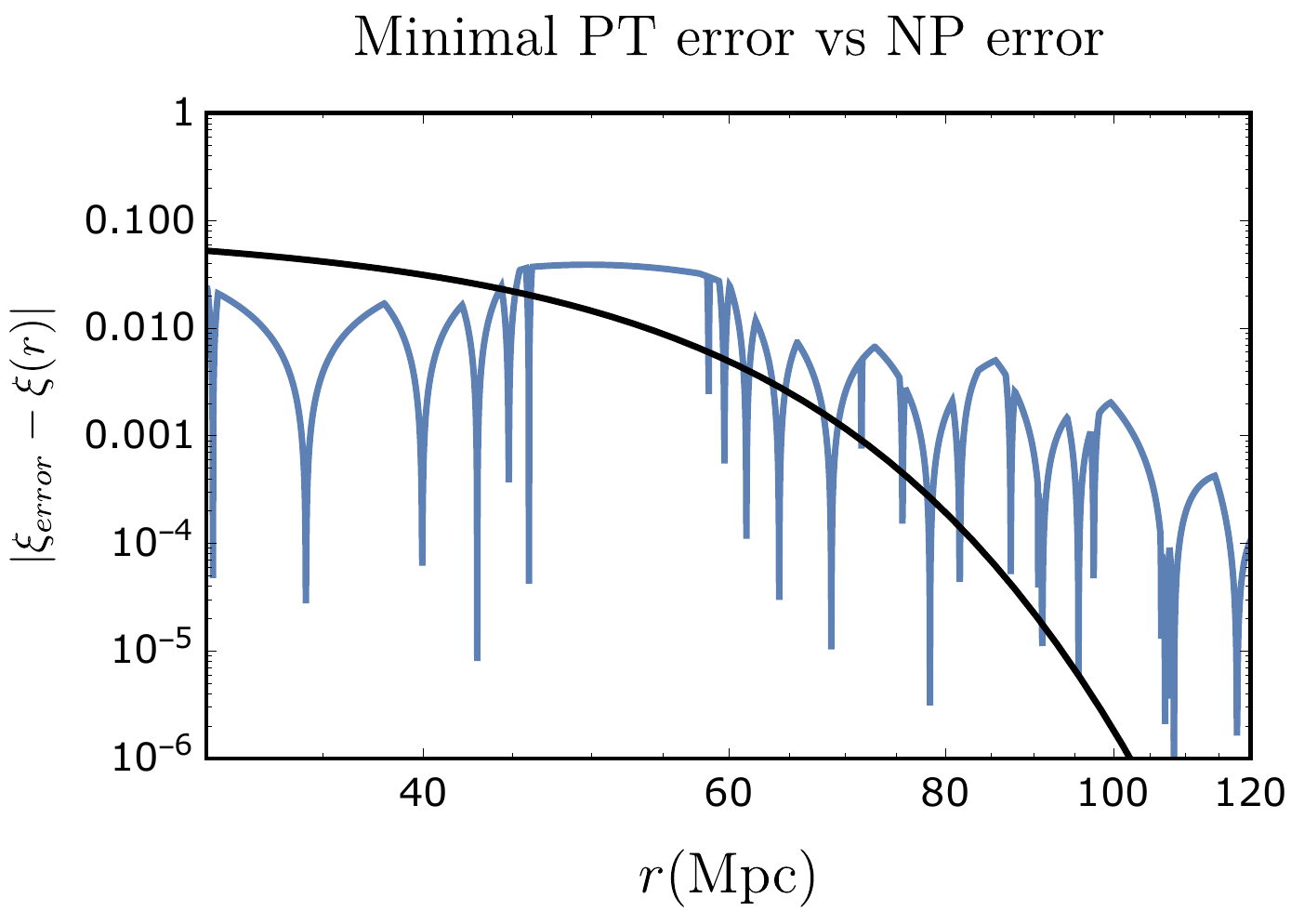}}
	\subfigure{\includegraphics[width=0.8\columnwidth]{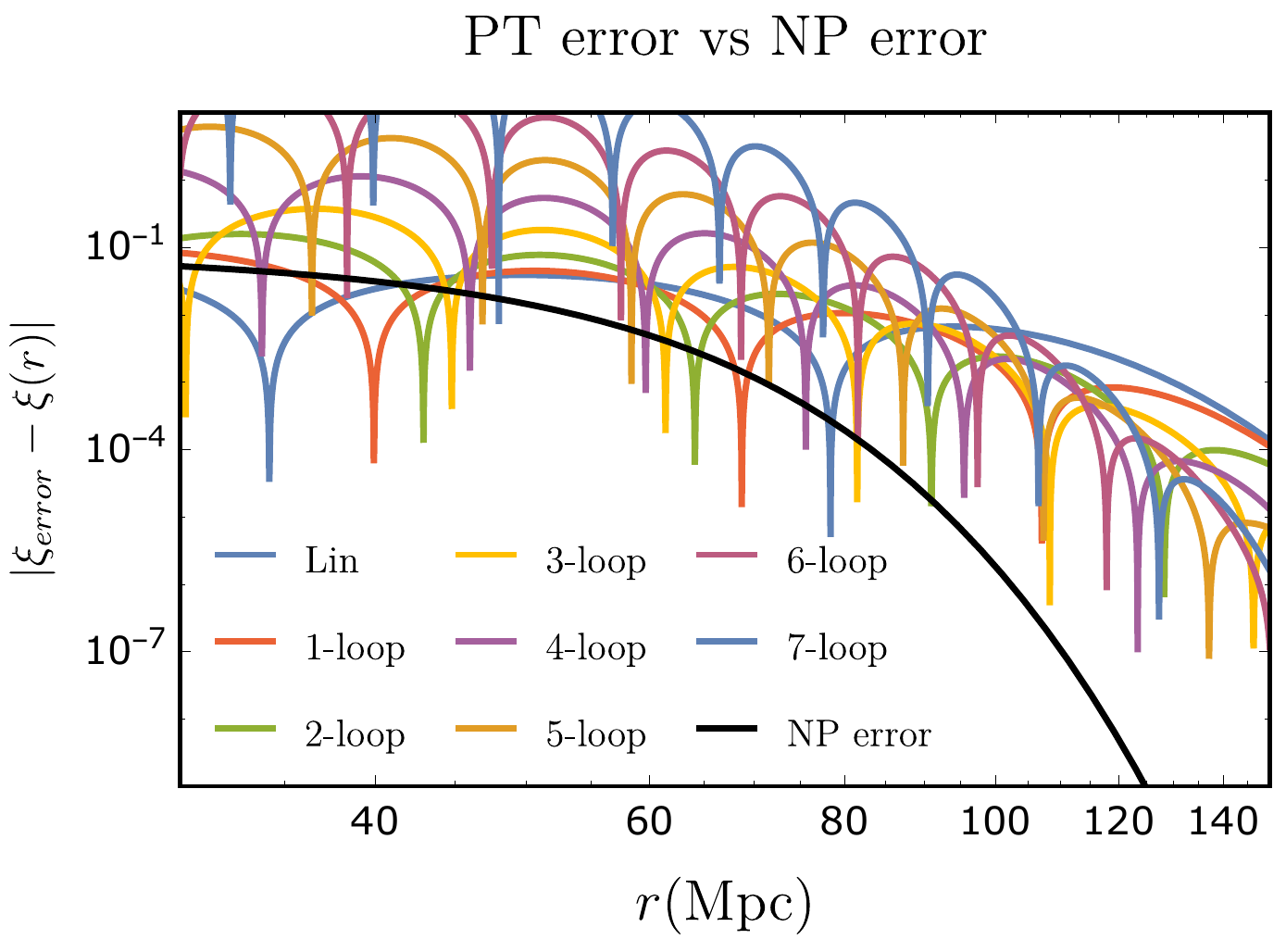}}
	\caption{The upper plot shows the minimal error one can get from the asymptotic perturbation theory by choosing the order that best approaches the exact result. The lower panel shows that for the scales plotted, the error indeed starts growing beyond 6 loops, so that we can trust the minimal error from this plot on these scales. The estimated NP error is smaller than the actual error on scales $  r>45  $ Mpc, but within a factor 10 on mildly nonlinear scales ($50<R<80$). Our estimate is orders of magnitude smaller than the actual error on larger scales. This is due to the relatively large polynomials in the integrand (see \eqref{nperror}. \label{nperrorplot1}}
\end{figure}

\begin{figure}[h!]
	\centering
	\includegraphics[width=\columnwidth]{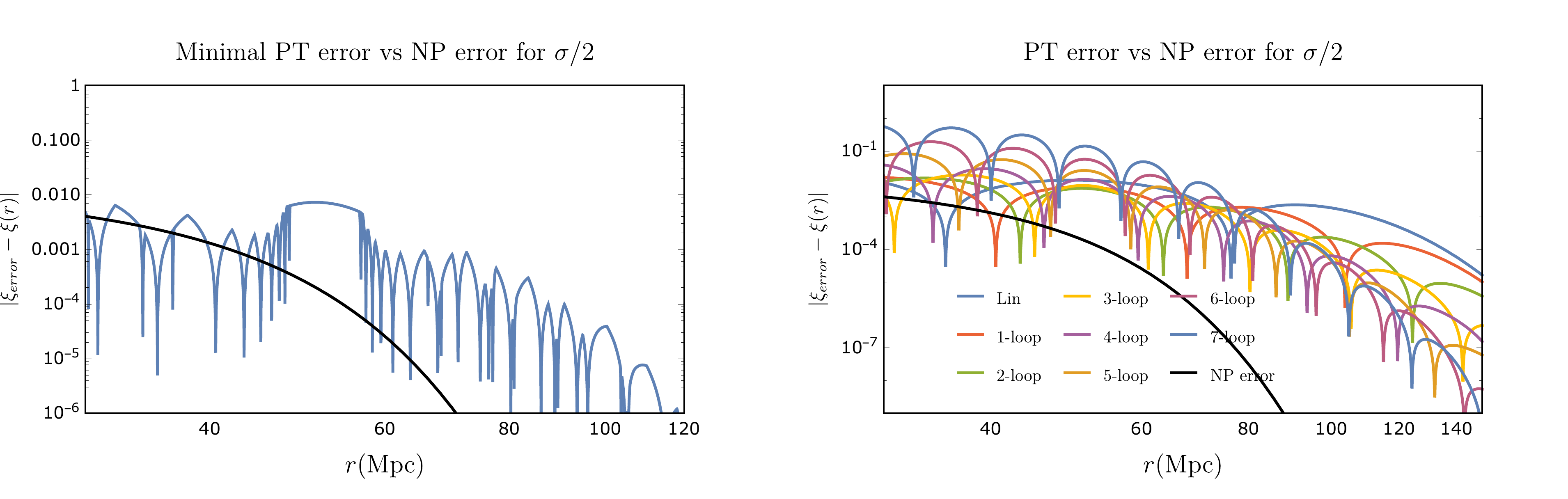}
	\caption{This is the same plot as Figure \ref{nperrorplot1}, but for which the initial linear variance is divided by a factor 2. It shows that the error estimate in fact worsens for smaller linear variance.\label{2nperrorplot}}
\end{figure}

\begin{figure}[h!]
	\centering
	\includegraphics[scale=0.8]{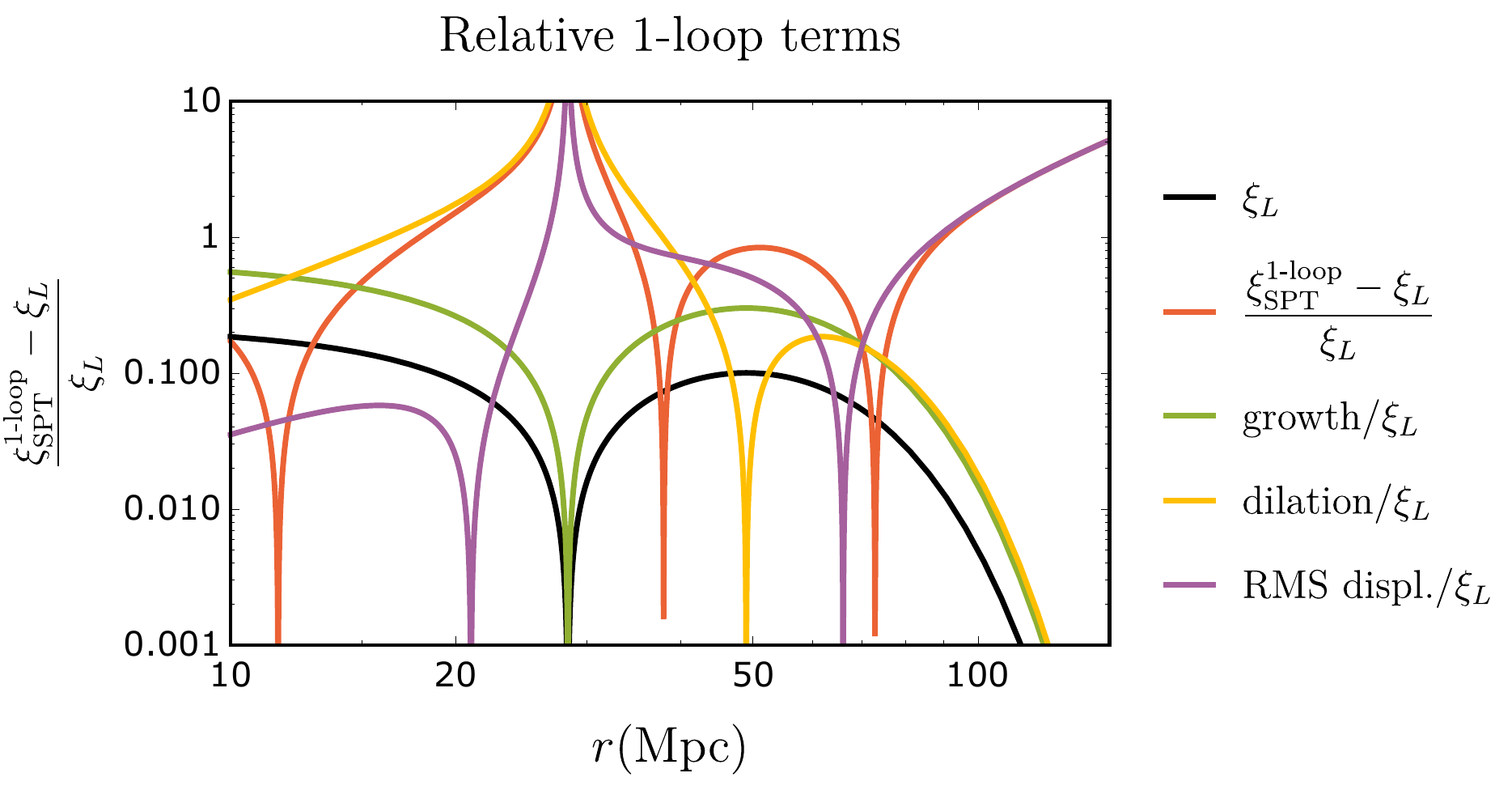}
	\caption{The plot shows the size of the various contributions \eqref{RMS} to $\xi_{\text{1-loop}}-\xi_{L}$ for our initial condition \eqref{exppk}. We plot the 1-loop terms devided by the linear term and compare to the linear term, since in a typical perturbation theory the 1-loop terms are of order $\xi_{L}^{2}$. On large scales this assumption crearly does not hold in this case. On mildly nonlinear scales the RMS displacement term is still one order of magnitude larger than naively expected. \label{zerolag}}
\end{figure}

Since PT is asymptotic for the correlation function, the optimal approximation is obtained at some $  n_{\text{opt}} $, beyond which the series starts diverging away from the true answer. We show how this optimal appriximation can be obtained numerically in Figure \ref{nperrorplot1}. We plot the PT errors, $  \xi^{\text{PT}}-\xi $, for increasing PT orders as function of distance. We also compare this error to a rough estimate (black thick line) obtained from the dimensionless integrand of the tail of the distribution
\begin{align}
\text{NP error} \sim \frac{1}{\sqrt{2\pi \tilde{\sigma}^{2}}}e^{-\frac{1}{2\tilde{\sigma}^{2}}},
\end{align}
where we used that $\lambda^{2}g_{r}(1)= \sigma^{2}(r)/r^{2}\equiv\tilde{\sigma}^{2}$. From Figure \ref{nperrorplot1} we see that this estimate of the non-perturbative error increasingly underestimates the actual error for $r>45Mpc$, presumably because of the large contribution from the polynomial term $  h $.  
To highlight the poor performance of perturbation theory, we reproduce Figure \ref{nperrorplot1} for smaller linear variance. In Figure \ref{2nperrorplot} we reduced our initial condition \eqref{exppk} by a factor 2. This shows that reducing the variance in fact worsens the error estimate on mildly nonlinear scales.  
We believe the reason for this poor performance of perturbation theory is twofold. First, the asymptotic nature of the series beyond the radius of convergence leads to relatively large polynomials $h$, which affects the integral significantly. Second, we note that the so called RMS displacement term, which is one of the one loop terms, is relatively large for our initial condition \eqref{exppk}. Using the notation and decomposition of \cite{McQuinn:2015tva}, the SPT 1-loop correlation function can be written as\footnote{Correcting the sign of the dilation term in (4.4) of \cite{McQuinn:2015tva}.}
\begin{align}\label{RMS}
\xi_{\text{SPT}}^{\text{1-loop}}=\xi_{L}(r)+\overbrace{3\xi_{L}^{2}(r)}^{\text{growth}}-\overbrace{4\xi_{L}^{\prime}(x)\int_{r}^{\infty}dx\,\xi_{L}(x)}^{\text{dilation}}+\overbrace{\frac{\sigma^{2}(r)}{2}\xi_{L}^{\prime\prime}(x)}^{\text{RMS displ.}}.
\end{align} 
The reason we have such a large RMS displacement, despite the absence of a BAO feature in our initial conditions is the fact that the linear variances are not simple power laws either, due to the exponential cutoff. This introduces another scales in the problem, which turns out to be picked up by the RMS displacement term. The large effect on large scales is illustrated in Figure \ref{zerolag}.

Finally, we stress that in all cases at least one of the tail contributions signals new physics, meaning they are probably hard to overcome analytically. For the PDF, the overdensity tail comes from the fact that large overdensities collapse, which is not captured by PT. Hence, a proper treatment of the tails requires an proper understanding of the collapsing process, which typically requires physics beyond the fluid approximation, such as a halo model, see appendix \ref{halo} for references. Similarly, the essential singularity in the integrand of the expression for the correlation function \eqref{corrfnc2} occurs when $q=\sigma_{\infty}x+r\leq0$, which can be interpreted as the contribution to the correlation function at distance $r$ from particles that were initially separated by distance $q\leq 0$, meaning it computes the contribution of particles that have crossed paths. These are beyond the reach of the fluid description and are also not properly captured by the ZA. A recent attempt to go beyond this limitation can be found in \cite{McDonald:2017ths}.

\subsection{Non-perturbative effects in Fourier space}

Here we note that the relation between non-perturbative effects in real and Fourier space is subtle. In particular, we investigate the analyticity in both cases. The results described here are a simple application of Paley-Wiener theorems. Suppose there is some error contribution to the correlation function 
\begin{align}
\xi(r)=\xi_{SPT}(r)+\xi_{NP}(r),
\end{align} 
whose asymptotic behavior is 
\begin{align}
\lim_{r\rightarrow\infty}\xi_{NP}(r)=e^{-\frac{1}{\sigma_{R}^{2}(r)}},
\end{align}
where $\sigma_{R}^{2}$ is the dimensionless variance. Now also suppose (realistically) that for large $r$, $\sigma_{R}^{2}(r)\sim r^{-n}$, for $n>1$. Then we can prove that such a term leads to an analytic contribution to the power spectrum for the following reason. Because of linearity 
\begin{align}
P(k)\supset \int_{r}e^{ikr}\xi_{NP}(r)=\int_{r}\sum_{n}\frac{1}{n!}(ikr)^{n}\xi_{NP}(r).
\end{align}
In order to use Fubini-Tonelli, we observe that
\begin{align}
\int_{r}\sum_{n}|\frac{1}{n!}(ikr)^{n}\xi_{NP}(r)|= \int_{r}\sum_{n}\frac{1}{n!}(kr)^{n}|\xi_{NP}(r)|=\int_{r}e^{kr}|\xi_{NP}(r)|<\infty,
\end{align}
by our assumption on the behavior of $\xi_{NP}(r)$ for large $r$. Thus we conclude that 
\begin{align}
P(k)\supset\int_{r}e^{ikr}\xi_{NP}(r)=\sum_{n}\frac{1}{n!}(ik)^{n}\int_{r}(r)^{n}\xi_{NP}(r),
\end{align}
meaning it is perfectly analytic in $k$. 

Conversely, if we are looking for errors that lead to non-analytic in $k$ type functions in Fourier space of the form $e^{-1/k^{n}}$ times some polynomial, we need all derivatives with respect to $k$ to vanish at $  k=0 $. Assuming that the moments of the correlation function exist, this means that the real space equivalent of this correction has vanishing moments
\begin{align}
\forall n\geq 0, \quad \int_{r}r^{n}\xi_{Error}(r)=0.
\end{align}
Such functions exist but, as shown above, cannot fall off at infinity too rapidly\footnote{Even though the examples we studied above all led to non-perturbative corrections to the correlation function that fall off exponentially at large $r$, it is not hard to understand that non-perturbative physics can alter the correlation function in a manner that does not fall off exponentially at large $r$ as well. As an example, consider just putting a rectangular bump of width $dk$ around $k_{nl}$ in the power spectrum: $P(k)=P_{old}(k)+Rect(k_{nl},dk)$. The Fourier transform of the latter term is $\propto\cos(k_{nl}r)\cos(r dk)/(rdk)$. Then it is not inconceivable that non-perturbative physics, which displaces particles around the nonlinear scale, completely changes the size of this bump in a non-calculable way. This therefore changes the respective term in the correlation function, which does not fall off exponentially at infinity. Moreover, it is an example of a contribution to the power spectrum that is non-analytic in $k$ (and which is not multiplied by $k^{4}$). More generally, we expect smooth, but non-analytic at $k=0$, approximations to the bump function to have similar behavior.}. We summarize these statements in Table \ref{table1}. 
\begin{table}
\begin{center}
\begin{tabular}{| p{4cm} | c | p{4cm} |}
	\hline 
	Real space    &        vs      & Fourier space                                    \\ \hline \hline
	Faster than $e^{-r}$ as $r\to\infty$ &  $\implies$            & Analytic in $k$ \\ \hline
	Zero moment functions (less than exponential, but more than polynomial fall off at infinity)   &  $\Longleftrightarrow$           & Vanishing derivatives at $k=0$ (the right to left implication makes sense for interchangeable derivative and integral only)                           \\ \hline
\end{tabular}
\end{center}
\caption[Table caption text]{Relation between asymptotic behavior of functions in real space and analyticity in Fourier space.}
\label{table1}
\end{table}


\subsection{Relation non-pertubative error and EFT-terms}\label{EFT}

One might wonder how the effects beyond perturbation theory that we discuss in this paper are related to short scale non-perturbative physics that is captured by the EFT of LSS. The fact that they are qualitatively different can be understood as follows. Observe that the non-perturbative effects we are discussing exist even if we cut off the power spectrum on short scales, such that there is zero power left. In that case, counterterms become irrelevant, but the effect we describe remains. Physically, the EFT is designed to capture the effect of unknown, short scale physics on long scales. The effect we are describing, however, are in a sense intrinsic limitations of the long scale physics \textit{itself}. In the first place, the mere asymptotic nature of perturbation theory can put a limit on the level of precision with which we can describe long scale physics, even in the absence of any coupling to short modes. Moreover, we argued that since long scale observables are statistical in nature, they always rely on a `tail contribution' (rare events) which are beyond perturbative understanding even on large scales. Nonetheless, since in a $\Lambda$CDM universe the EFT contributions on mildly nonlinear scales appear to be much larger than the non-perturbative effects we are referring to in this paper, the EFT of LSS is still very useful \cite{Carrasco:2012cv}.  


\section{Conclusion and outlook}\label{sec:6}

In this paper we made progress towards a better understanding of the reach of perturbation theory for large scale structures and the relevance of non-perturbative (NP) effects. In the context of 1D physics, we proved the asymptotic nature of perturbation theory for the real space correlation function and count-in-cell cumulants on all scales. In both cases, the proof is based on the presence of an essential singularity in the domain of the defining integral. Interestingly, this singularity has physical significance, signaling the collapse of cells or multi-streaming events. This adds to the intuition about NP effects in the halo model, which we review in Appendix \ref{halo}. Altogether, \textit{this suggests that there is indeed a floor to the reach of perturbative approaches to LSS based on fluid dynamics}. For our initial condition \eqref{exppk}, we found that in real space the best possible perturbative approximation is typically worse than a naive guess for the error based on the size of the tail of the integrand \eqref{nperror},
\begin{align}
\xi_{\text{NP}}(r) \sim \frac{1}{\sqrt{2\pi }} \frac{r}{\sigma(r)}e^{-\frac{r^{2}}{2{\sigma(r)}^{2} }} \quad \text{(naive estimate)},
\end{align}
where $r/\sigma$ is dimensionless. It would be interesting to study the implication for these non-perturbative effects for the position and shape of the BAO feature. 

Conversely, we found that 1D SPT is convergent for the Fourier space observables we considered: the power spectrum for both Gaussian and non-Gaussian initial conditions, and the bispectrum for Gaussian initial conditions (assuming $  \Lambda $CDM-like initial conditions). We expect this to extend to other Fourier space observables in 1D as well. However, we do not expect this to extend to 3D. The reach of perturbation theory in 3D was investigated in for instance \cite{Blas:2013aba} and \cite{Baldauf:2015tla}, but a proof of the asymptotic nature of perturbation theory in 3D remains an open problem. 

We stress that, even in 1D, our findings do not yet imply that perturbation theory converges to the \textit{physically} correct answer. In fact, even neglecting EFT corrections, we expect non-perturbative effects of the form
\begin{align}
\text{NP error power spectrum} \sim \frac{1}{\sqrt{2\pi\Delta^{2}(k)}}e^{-\frac{1}{2\Delta^{2}(k)}},
\end{align}
where $\Delta$ is the dimensionless Fourier space variance, to play a role for the power spectrum as well. Again, the intuitive reason is the statistical nature of cosmological correlators. This means that even on large scales there are always some rare events for which the dimensionless quantity $k\delta(k)$ is larger than unity and the fluid description fails. The inevitable statistical contribution of these rare events to cosmological correlators is then expected to be exponentially suppressed. An analytic or quantitative understanding of this statement for Fourier space observables remains an open problem. 

The importance of a solid control over theoretical errors in predictions for LSS observables was argued in \cite{Baldauf:2016sjb,Welling:2016dng}. One of the hopes would therefore be to provide some sort of fitting function for NP effects on mildly nonlinear scales. Unfortunately, it is not clear a priori if there are any symmetries or other physical arguments to find such a function, despite significant effort in this direction in the literature. Here, we presented a novel construction for the count-in-cell PDF in 1D, which could be useful for the construction of its 3D analogue. However, a thorough understanding of theoretical errors for the 3D PDF seems to still be lacking.

Finally, the most useful step going forward is probably a combined effort of both analytical and numerical searches for NP effects, as advocated in \cite{Baldauf:2015tla}.


\appendix

 
\section*{Aknowledgements} We would like to thank Simon Foreman, Uro\v s Seljak, Zvonimir Vlah, Martin White, Yvette Welling and Cora Uhlemann for useful discussions. We would also like to thank Garrett Goon and John Stout for comments on an earlier version of the paper. We are particularly thankful to Simon Foreman for sharing with us his results on scaling universes and to Gabriele Trevisan for collaboration on deriving the results of subsections \ref{Gab1} and \ref{Gab2}. Finally, we would like to thank the anonymous referee for significant and constructive recommendations. E.P. is supported by the Delta-ITP consortium, a program of the Netherlands organization for scientific research (NWO) that is funded by the Dutch Ministry of Education, Culture and Science (OCW). 


\section{Planar collapse}\label{planar}

Here we show that the Lagrangian equation \eqref{Leom} is also found for the evolution of the density in finite cells. We derive this within Newtonian cosmology. Let us consider the evolution of a cylinder, whose symmetry axis is along the $x$-direction with physical radius $a$ and length $R$. In the Newtonian approximation, we can use Gauss' law in physical coordinates
\begin{align}
\int_{V}\nabla^{2}\phi=\int_{S}\left(\vec{\nabla}\phi\right) \cdot \vec{n},
\end{align}
where $V$ is the volume of the cell, $S$ is the surface, and $\vec{n}$ is the surface normal vector. Using the Poisson equation,
\begin{align}
\nabla^{2}\phi=4\pi G \rho(x),
\end{align}
this gives
\begin{align}
4 \pi G M=\int_{S}\left(\vec{\nabla}\phi\right) \cdot \vec{n},
\end{align} 
where and $M$ is the total mass inside the cylinder. We now wish to evaluate the surface integral. In order to compute the flux through the side of the cylinder, observe that shuffeling matter in the $x$-direction cannot induce a relative force between particles in the orthogonal directions. The flux through the side of the cylinder is therefore equal to what it is in the completely homogeneous case, and its motion is according to the average expansion of the universe. We can therefore interpret $a$ as the scale factor, with corresponding dynamics. The cylinder surface integral then becomes
\begin{align}
\int_{S}\left(\vec{\nabla}\phi\right) \cdot \vec{n}&=2\pi a R \left(\nabla_{a} \phi\right)+\pi a^{2}\left(\nabla_{x} \phi(x_{1}+R)-\nabla_{x} \phi(x_{1})\right) \nonumber \\
&=-2\pi a R \ddot{a}-\pi a^{2}\ddot{R}, 
\end{align}  
where $x_{1}$ should be considered as a label for this cell, tracking the left outer edge. We define the average density in this cell as
\begin{align}
\delta_{R}(x_{1})=\frac{1}{R}\int_{x_{1}}^{x_{1}+R}\delta(x)dx.
\end{align}
Mass conservation inside the cell then relates the average density to the size of the cell,
\begin{align}
R=\frac{M}{\pi a^{2}\bar{\rho}(1+\delta_{R})}.
\end{align}
Plugging this into the integrated Poisson equation
\begin{align}
4 \pi G M=-2\pi a R \ddot{a}-\pi a^{2}\ddot{R},
\end{align}
and using the Friedmann equations for a matter dominated universe,
\begin{align}
\begin{cases}
H=\frac{\dot{a}}{a}=\frac{8\pi G}{3}\bar{\rho} \\
\dot{H}+H^{2}=\frac{\ddot{a}}{a}=-\frac{4\pi G}{3}\bar{\rho},
\end{cases}
\end{align}
this yields the following equation for the evolution of the average density:
\begin{align}\label{1Dcollapse}
\ddot{\delta}_{R}+2H\dot{\delta}_{R}-2\frac{\dot{\delta}_{R}^{2}}{1+\delta_{R}}=4\pi G\bar{\rho}\delta_{R}(1+\delta_{R}).
\end{align}
As noted before, the solution to this equation is found from the nonlinear transformation \eqref{1Dsolution}.

 
\section{Growing and decaying modes in spherical collapse}\label{app:12}

The growing and decaying modes in the spherical collapse solution are obtained by taking $\eta$ and $C$, and therefore $t$, small. Let us first define
\begin{align}
y\equiv \frac{t-C}{B}=\eta-\sin\eta.
\end{align} 
Then, in the small $\eta$ limit, we find to second order
\begin{align}
y=\frac{\eta^{3}}{6}-\frac{\eta^{5}}{120}.
\end{align}
This we can invert perturbatively to find to second order
\begin{align}
\eta=\left(6y+\frac{(6y)^{2/3}}{20}\right)^{1/3}=(6y)^{1/3}+\frac{y}{10}.
\end{align}
Expanding the denominator in \eqref{deltaSC}, to second order we find
\begin{align}
\delta=\frac{9t^{2}}{2B^{2}}\left(\frac{8}{\eta^{6}}+\frac{2}{\eta^{4}} \right)-1.
\end{align}
Plugging in the expression for $\eta$ in terms of $y$, the second order expression becomes
\begin{align}
\delta=\frac{9t^{2}}{2B^{2}}\left(\frac{2}{9y^{2}}+\frac{1}{5y(6y)^{1/3}} \right)-1.
\end{align}
Finally, we substitute back the expression for $y$ in terms of $t$ and $C$, and expand to first order in $C$, thereby assuming $C\ll t \ll 1$, which yields
\begin{align}
\delta_{L}=\frac{3}{10}\left(\frac{9}{2}\right)^{1/3}\left(\frac{t}{B}\right)^{2/3}+\frac{2C}{t}.
\end{align}


\section{Displacement field symmetries and the bispectrum}\label{ssec:tt}

The equation of motion for the stochastic field $\Psi$, \eqref{eqpsi}, is invariant under two separate symmetries
\be
Q_{1}&:&\quad \Psi(q)\rightarrow\psi'(q)\equiv\Psi(q+c_{1}(t))\,,\\
Q_{2}&:&\quad \Psi(q)\rightarrow\psi'(q)\equiv\Psi(q)+c_{2}(t)\,,
\ee
for arbitrary functions of time $  c_{i}(t) $. It is straightforward to see that both these transformations correspond to translations of $  \delta $, namely (in $  d $-dimensions)
\be\label{ZeldR}
\delta(x)=\int d^{d}q \delta_{D}\left(  x-q-\Psi(q)\right)\rightarrow \delta'(x)\equiv \delta(x+c_{1}-c_{2})\,.
\ee
Therefore, for $c_{1}=c_{2}=c$, the linear combination $  Q_{+}\equiv Q_{1}+Q_{2} $ leaves $  \delta $ exactly invariant (not covariant), while the orthogonal combination $   Q_{-}\equiv Q_{1}-Q_{2} $ induces a translation $  \delta(x)\rightarrow \delta(x+2c) $. 

Now there are two different situations:
\begin{itemize}
	\item \textit{If and only if we are interested in correlators of $  \delta $}, we can perform a $  Q_{+} $ transformation without changing the value of the $  \delta $ correlator. Also, assuming statistical homogeneity, we can also perform a $   Q_{-}  $ transformation, without changing the value of the $  \delta $ correlator. Hence, we use both $  Q_{1} $ and $  Q_{2} $ at will to simplify our $  \delta $ correlators.
	\item \textit{If instead we also care about correlators of $  \Psi $ that do not combine into a $  \delta $}, things are trickier. In principle one can invert \eqref{ZeldR} exactly getting $  \Psi(\delta) $ and then again both $  Q_{+} $ and $  Q_{-} $ annihilate correlators. In practice though we are not able to invert \eqref{ZeldR} but only at linear order, which is a good approximation if the initial perturbations are small $  |\Psi|,|\delta|\ll1 $. This linearized inversion, \eqref{linearized}, assumes $  |\psi|\ll 1 $ and therefore breaks $  Q_{2} $, but not $  Q_{1} $. So, when we compute $  \Psi $ correlators using this approximation, or equivalently declaring the initial displacement power spectrum \eqref{notdelta}, we can perform $  Q_{1} $ without changing the value of the correlator by invoking statistical homogeneity, but we cannot perform $  Q_{2} $ transformations.
\end{itemize}

Let us move on to compute the power spectrum, but inverting the usual definition 
\be
\ex{\delta(\vec{k})\delta(\vec{k}')}\equiv \left( 2\pi \right)^{d} \delta_{D}^{(d)}\left(  \vec{k}+\vec{k'}\right)P(k)\,,
\ee
namely
\be
P(k)=\int_{\vec{k}'}\ex{\delta(\vec{k})\delta(\vec{k}')}\,.
\ee
Using the Fourier transform of \eqref{ZeldR}, namely
\be\label{Zeld}
\delta(k)=\int dq e^{-ik\left[ q+\Psi(q) \right]}\,,
\ee
one finds three types of terms
\be
P(k)=\int_{k'qq'}e^{-i\left(  kq+k'q'\right)}\left[  \ex{e^{-i\left( k\Psi +k'\Psi'\right)}}-\ex{ e^{-ik\Psi}}-\ex{e^{-ik\Psi} }+1\right]\,.
\ee
Using the symmetry $  Q_{2} $
\be
\ex{ e^{-ik\Psi}}=\ex{ e^{-ik\left( \Psi -\Psi\right)}}=1\,.
\ee
One might be slightly worried about choosing the parameter $  c(t) $ of the $  Q_{2} $ transformation to be a random variable, rather than just a function. This can in principle be justified by looking at the PDF formulation of the correlators, but we won't do it here. Using both $  Q_{1} $ and $  Q_{2} $ we can re-write
\be
\ex{e^{-i\left( k\Psi +k'\Psi'\right)}}= \ex{e^{-i k\left( \Psi(q-q')-\Psi(0)  \right)-i k'\left( \Psi(0)-\Psi(0) \right)} }=\ex{e^{-ik\Delta \Psi(q-q')}}\,,
\ee
with
\be
\Delta \Psi(q)\equiv \Psi(q)-\Psi(0)\,.
\ee
These two tricks lead to the standard result
\be
P(k)&=&\int dq  \, e^{-ikq}\left[  \ex{e^{-ik\Delta \Psi(q)}}-1\right]\,.
\ee
The calculation of the bispectrum proceeds very similarly starting from the definition
\be
B(k_{1},k_{2})\equiv \int_{k_{3}} \ex{\delta(k_{1})\delta(k_{2})\delta(k_{3})}\,.
\ee
The result is the same as one would have guessed from the get go, \eqref{bispectrum}.


\section{Convergence of SPT for the Bispectrum}\label{app:bispectrum}

The bispectrum in 1D can be intuitively expressed as
\begin{align}\label{bispectrum}
B(k_{1},k_{2})=V^{-1}\langle|\delta(k_{1})\delta(k_{2})\delta(-k_{1}-k_{2})|\rangle,
\end{align}
where $V$ is the 1D spatial volume. In Appendix \ref{ssec:tt}, we derive this expression more formally from symmetries. Plugging in the ZA expressions for the density, we obtain
\begin{align}
B(k_{1},k_{2})=V^{-1}\langle\int & dq_{123}e^{-i k_{1}q_{1}}e^{-i k_{2}q_{2}}e^{i (k_{1}+k_{2})q_{3}}\nonumber\\
&\times\left(e^{-ik_{1}\Psi_{1}}-1\right)\left(e^{-ik_{2}\Psi_{2}}-1\right)\left(e^{i(k_{1}+k_{2})\Psi_{3}}-1\right)\rangle,
\end{align}
where $\Psi_{i}\equiv\Psi(q_{i})$. To guide our intuition, we can already observe that this expression goes to zero as any of the displacements goes to zero. Working out the terms in brackets, we find
\begin{align}\label{B1}
B(k_{1},k_{2})=V^{-1}\int & dq_{123}e^{i k_{1}(q_{3}-q_{1})}e^{i k_{2}(q_{3}-q_{2})}\times \nonumber \\
&\Bigg[\langle e^{i k_{1}(\Psi_{3}-\Psi_{1})}e^{i k_{2}(\Psi_{3}-\Psi_{2})}\rangle-\langle e^{-i k_{1}\Psi_{1}}e^{-i k_{2}\Psi_{2}}\rangle \nonumber \\
&-\langle e^{-i k_{1}(\Psi_{1}-\Psi_{3})}e^{-i k_{2}\Psi_{3}}\rangle+\langle e^{-ik_{1}\Psi_{1}}\rangle \nonumber \\
&-\langle e^{-i k_{2}(\Psi_{2}-\Psi_{3})}e^{-i k_{1}\Psi_{3}}\rangle+\langle e^{-ik_{2}\Psi_{2}}\rangle \nonumber \\
&-\langle e^{-i (k_{1}+k_{2})\Psi_{3}}\rangle-1 \Bigg].
\end{align}
Now it's just a matter of correlating. For this we use the following observations. First, we use the cumulant expansion theorem, which is easily proved in this context. Namely, all expectation values are of the form 
\begin{align}
\langle e^{i\vec{\Psi}\cdot\vec{a}}\rangle = \int d\Psi_{1}d\Psi_{2}d\Psi_{3}e^{i\vec{\Psi}\cdot\vec{a}}\frac{1}{\sqrt{\det{2\pi C}}}e^{\frac{1}{2}\vec{\Psi}^{T}C^{-1}\vec{\Psi}},
\end{align}
where $\vec{\Psi}^{T}=\left(\Psi_{1},\Psi_{2},\Psi_{3}\right)$, $\vec{a}$ is some $\Psi$-independent vector, and $C$ the multi-variate Gaussian correlation matrix. Completing the square in the exponent then allows us to evaluate the Gaussian integral, and we are left with 
\begin{align}
\langle e^{i\vec{\Psi}\cdot\vec{a}}\rangle = e^{-\frac{1}{2}\vec{a}^{T}C\vec{a}}.
\end{align}
Finally, we observe that 
\begin{align}
\vec{a}^{T}C\vec{a}=\langle\left(\vec{a}\cdot\vec{\Psi}\right)^{2}\rangle,
\end{align}
proving the cumulant expansion theorem in this case. Evaluating the two point correlations, we write
\begin{align}
2\langle\Psi(q_{1})\Psi(q_{2})\rangle=\sigma_{q}^{2}(q_{1}-q_{2})=-\sigma^{2}(q_{1}-q_{2})+\sigma_{\infty}^{2},
\end{align} 
with $\sigma_{\infty}^{2}=2\langle\Psi(q_{1})\Psi(q_{1})\rangle$. Note that this means the three expectation values in \eqref{B1} that only depend on a single coordinate evaluate to a constant. They can therefore be taken out of the integral and the remaining integral can then be computed using a change of coordinates to yield (up to potential factors of $2\pi$)
\begin{align}
V^{-1}\int dq_{123}e^{i k_{1}(q_{3}-q_{1})}e^{i k_{2}(q_{3}-q_{2})}=V^{-1}\delta_{D}(0)\delta_{D}(k_{1})\delta_{D}(k_{2})=\delta_{D}(k_{1})\delta_{D}(k_{2}).
\end{align}
Because of these $\delta$-functions,  we can set the momenta of the single-coordinate-expectation-values to zero at the start, and replace the full integral with unity. Using similar changes of coordinates, and using the same trick we used in \eqref{PStrick1} and \eqref{PStrick2}, we finally find the following expression for the bispectrum
\begin{align}
& B(k_{1},k_{2})=V^{-1}\int  dq_{123}e^{i k_{1}(q_{3}-q_{1})}e^{i k_{2}(q_{3}-q_{2})}\times \nonumber \\
&\Bigg[e^{-\frac{1}{2}\left[k_{1}^{2}\sigma^{2}(q_{1}-q_{3})+k_{2}^{2}\sigma^{2}(q_{2}-q_{3})+2k_{1}k_{2}\left(\sigma_{\infty}^{2}-\sigma_{q}^{2}(q_{1}-q_{3})-\sigma_{q}^{2}(q_{2}-q_{3})+\sigma_{q}^{2}(q_{1}-q_{2})\right)\right]} \nonumber\\
&-e^{-\frac{1}{2}\left[k_{1}^{2}\sigma_{\infty}^{2}+k_{2}^{2}\sigma_{\infty}^{2}+2k_{1}k_{2}\sigma_{q}^{2}(q_{1}-q_{2})\right]} -e^{-\frac{1}{2}\left[k_{1}^{2}\sigma^{2}(q_{1}-q_{3})+k_{2}^{2}\sigma_{\infty}^{2}-2k_{1}k_{2}\left(\sigma_{q}^{2}(q_{1}-q_{3})-\sigma_{\infty}^{2}\right)\right]}\nonumber \\
&-e^{-\frac{1}{2}\left[k_{2}^{2}\sigma^{2}(q_{2}-q_{3})+k_{1}^{2}\sigma_{\infty}^{2}-2k_{1}k_{2}\left(\sigma_{q}^{2}(q_{2}-q_{3})-\sigma_{\infty}^{2}\right)\right]}  +2 \Bigg].
\end{align} 
Changing coordinates to $\tilde{q}_{1}=q_{1}-q_{3}$, $\tilde{q}_{2}=q_{2}-q_{3}$, $\tilde{q}_{3}=q_{3}$, and dropping the tildes (note that this transformation has unit determinant), we obtain
\begin{align}
& B(k_{1},k_{2})=\int  dq_{12}e^{-i k_{1}q_{1}}e^{-i k_{2}q_{2}}\times \nonumber \\
&\Bigg[e^{-\frac{1}{2}\left[k_{1}^{2}\sigma^{2}(q_{1})+k_{2}^{2}\sigma^{2}(q_{2})+2k_{1}k_{2}\left(\sigma_{\infty}^{2}-\sigma_{q}^{2}(q_{1})-\sigma_{q}^{2}(q_{2})+\sigma_{q}^{2}(q_{1}-q_{2})\right)\right]} \nonumber\\
&-e^{-\frac{1}{2}\left[k_{1}^{2}\sigma_{\infty}^{2}+k_{2}^{2}\sigma_{\infty}^{2}+2k_{1}k_{2}\sigma_{q}^{2}(q_{1}-q_{2})\right]} -e^{-\frac{1}{2}\left[k_{1}^{2}\sigma^{2}(q_{1})+k_{2}^{2}\sigma_{\infty}^{2}-2k_{1}k_{2}\left(\sigma_{q}^{2}(q_{1})-\sigma_{\infty}^{2}\right)\right]}\nonumber \\
&-e^{-\frac{1}{2}\left[k_{2}^{2}\sigma^{2}(q_{2})+k_{1}^{2}\sigma_{\infty}^{2}-2k_{1}k_{2}\left(\sigma_{q}^{2}(q_{2})-\sigma_{\infty}^{2}\right)\right]}  +2 \Bigg]. \label{refep}
\end{align} 
Let us first focus on the constant terms in the exponents. When one is interested in its contribution to the bispectrum, one is free to replace the constant 2 with any function of $k_{1}$ and $k_{2}$ that equals 2 when the $k_{i}$ are zero. In this case it is convenient to replace it with 
\begin{align}
2\to e^{k_{1}k_{2}\sigma_{\infty}^{2}} +e^{\frac{1}{2}(k_{1}^{2}+k_{2}^{2})\sigma_{\infty}^{2}}.
\end{align}
Schematically, the perturbative expansion of the integrand looks like
\begin{align}
\sum_{n}\frac{1}{n!}\left(\sum_{i}(f_{i}(q)+f_{i,\infty})^{n}\right),
\end{align}
where $\lim_{q\to\infty}q^{2}f_{i}(q)=0$, and $\sum_{i}f_{i,\infty}=0$ because of the $\delta$-function replacement above. The sum over $i$ denotes the five terms of the integrand. In order to use Fubini-Tonelli, we wish to bound 
\begin{align}
\sum_{n}\frac{1}{n!}\bigg|\sum_{i}^{i_{max}}(f_{i}(q)+f_{i,\infty})^{n}\bigg|.
\end{align}
Now, for $q$ large, by definition $f_{i}(q)<f_{i,\infty}$. Let $C=\max_{i}\{|f_{i,\infty}|\}$, and $g(q)=\max_{i}\{|f_{i}(q)|\}$. Then we can approximate the above sum as
\begin{align}
\sum_{n}\frac{1}{n!}\bigg|\sum_{i}^{i_{max}}(f_{i}(q)+f_{i,\infty})^{n}\bigg|\leq g(q)\sum_{n}\frac{1}{n!}i_{max}2^{n}C^{n-1}=g(q)i_{max}\frac{1}{C}e^{2C}.
\end{align}
Since this goes to zero faster than $q^{2}$ as $q_{i}\to \infty$, we conclude that
\begin{align}
\int  dq_{12}\sum_{n}\frac{1}{n!}\bigg|b_{n}(q_{1},q_{2})\bigg|<\infty,
\end{align}
where $b_{n}$ is the $n$-th order term in the expansion of the integrand. Since the absolute value of the oscillating factors is obviously also bounded, this argument holds for the bispectrum as well, such that we conclude that Fubini-Tonelli indeed applies.

In conclusion, we can exchange the sum with the integral in \eqref{refep} and this shows that the SPT bispectrum converges to the ZA bispectrum, in analogy with what happens for the power spectrum. 


\section{Convergence of SPT for the Power Spectrum for NG initial conditions}\label{app:ng}

Here we are interested in the convergence of the power spectrum for NG initial conditions. Of course, the answer to this question depends on the type of initial conditions. We specify the assumptions we make below, which hold for perturbative non-Gaussianity. Using the cumulant expansion theorem as in equation (2.27) of \cite{McQuinn:2015tva}, we find an expression of the form
\begin{align}
P^{NG}(k)=\int_{-\infty}^{\infty}dq e^{-ikq}\left[e^{-k^{2}\sigma^{2}(q)/2+ik^{3}\sigma_ {3}(q)/3!-k^{4}\sigma_{4}(q)/4!+\ldots}-1\right]\,,
\end{align}
where $  \sigma_{3,4,\dots} $ characterize the type of primordial non-Gaussianity. Let us now assume the sum in the exponent can be decomposed as $\sum_{i}f_{k,i}^{\infty}+f_{k,i}(q)$, such that $\lim_{q\to\infty}f_{k,i}(q)=0$. Here the $f_{k,i}(q)$ are obtained from the cumulants by subtracting their asymptotic values. For instance, 
\begin{align}
f_{k,4}(q)=-k^{4}\sigma_{4}(q)/4!-f_{k,4}^{\infty} \, ; \quad f_{k,4}^{\infty}=-\lim_{q\to\infty}k^{4}\sigma_{4}(q)/4! \, .
\end{align}
Once again, we can rewrite the 1 in the integrand as $e^{\sum_{i}f_{k,i}^{\infty}}$. Then 
\begin{align}
P^{NG}(k)=\int_{-\infty}^{\infty}dq e^{-ikq}e^{\sum_{i}f_{k,i}^{\infty}}\left[e^{\sum_{i}f_{k,i}(q)}-1\right].
\end{align}
One has to be a bit careful about the expansion, since clearly the $f_{k,i}$ count at different orders in PT. However, at every order in PT we get some sum of terms, whose absolute value is always smaller than the sum of the absolute values of the individual terms. Resumming these absolute values is easy. If 
\begin{align}
P^{NG}(k)=\int_{-\infty}^{\infty}dq \sum_{n}p^{NG}_{n}(k,q),
\end{align}  
we get
\begin{align}
\int_{-\infty}^{\infty}dq \sum_{n}|p^{NG}_{n}(k,q)|\leq e^{\sum_{i}|f_{k,i}^{\infty}|}\int_{-\infty}^{\infty}dq \left[e^{\sum_{i}|f_{k,i}(q)|}-1\right].
\end{align}
From this we easily conclude the condition for convergence of PT: if all $f_{k,i}(q)$ satisfy $\lim_{q\to\infty}q^{2}f_{k,i}(q)=0$ and $\sum_{i}|f_{k,i}^{\infty}|<\infty$, then the expression above converges. In this case, the Fubini-Tonelli theorem applies and SPT converges to ZA, even for non-Gaussian initial conditions.


\section{(non-)Convergence of RPT for the correlation function}\label{app:RPT}

As we have seen previously, the power spectrum admits a convergent SPT expansion. Let us now assume that the exact correlation function exists and is finite
\begin{align}
\xi(x)=\int_{k}e^{ikx}P(k)=\int_{k}e^{ikx}\sum_{n=1}^{\infty}P_{n}(k)<\infty\,.
\end{align}
Notice that, since $  P_{n}(k)=P_{n}(-k) $ for every $  n $, we can also use that the Fourier transform reduces to the cosine transform 
\begin{align}
\xi(x)=\int_{k}\cos \left( kx \right)P(k)=\int_{k}\cos \left( kx \right)\sum_{n=1}^{\infty}P_{n}(k)<\infty\,.
\end{align}
Then, a candidate perturbative expansion is given by
\begin{align}
\xi(x)\overset{?}{=}\sum_{n=1}^{\infty}\xi_{n}(x)=\sum_{n=1}^{\infty}\int_{k}e^{ikx}P_{n}(k)\,.
\end{align}

\begin{figure}[h]
	\centering
	\includegraphics[width=\columnwidth]{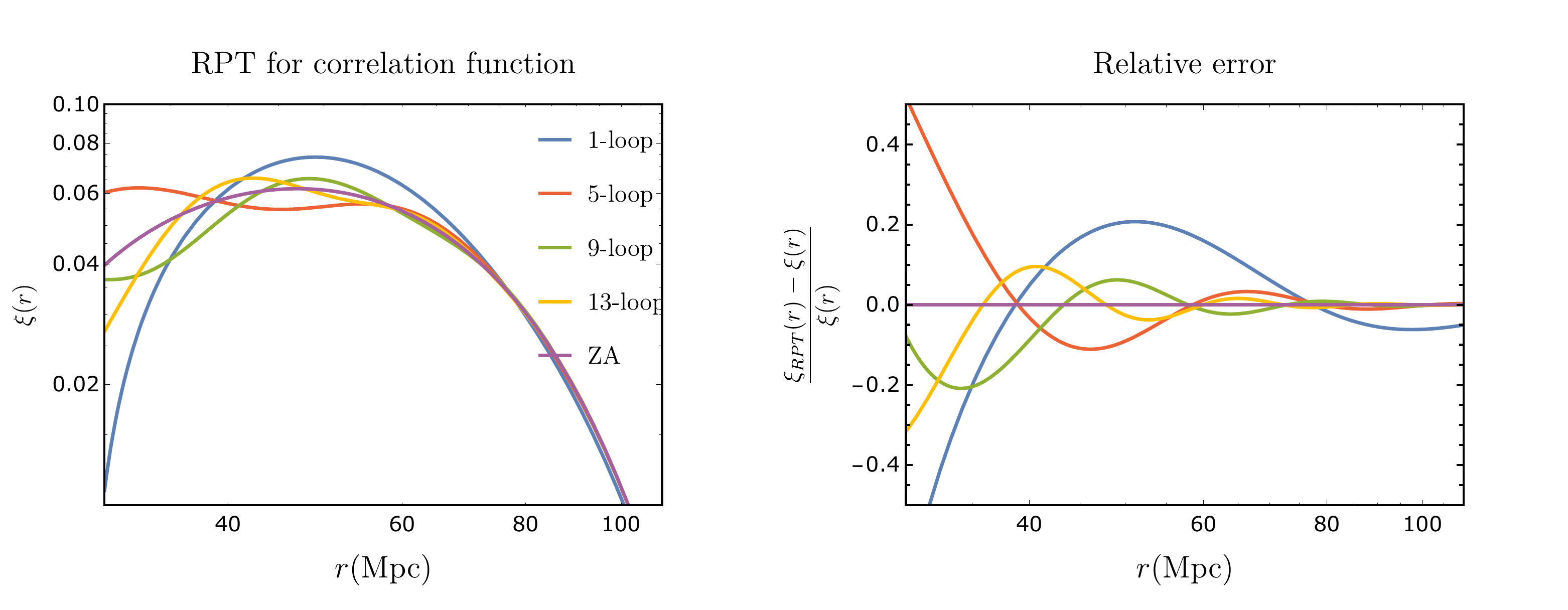}
	\caption{The plot shows the ZA correlation function and its RPT approximations. To the order we consider, there is no sign of divergence by increasing loop order, although the improvement from 9 to 13 loops is small. Lacking an analytical proof, we consider the (non-)convergence of RPT inconclusive. We use \eqref{exppk} as initial condition. \label{RPTplots}}
\end{figure}
This perturbative expansion converges if we are allowed to exchange the sum with the integrand. For SPT, we proved in subsection \ref{nonccf} that this is not the case. One obstacle was that $  P_{n}(x) $ are not all positive. To this end, let us consider Renormalized Perturbation Theory (RPT) \cite{Crocce:2005xy}. At least for LCDM\footnote{Note that there is a wrong minus sign in eq (2.39) of \cite{McQuinn:2015tva}}
\begin{align}
P_{n}^{\text{RPT}}(k)=e^{-\frac{k^{2}}{2}\ssi}\frac{1}{n!}\int_{q}e^{-ikq}\left[ \frac{k^{2}}{2}\ssq(q) \right]^{n}>0\,.
\end{align}
Then, indeed, all $P_{n}$ are positive, as shown in \cite{Crocce:2005xy}. Thus we can use Fubini-Tonelli theorem to invert the sum with the integral, provided that
\begin{align}
\int_{k}\sum_{n=1}^{\infty} |\cos \left( kx \right)P_{n}^{\text{RPT}}(k)|<\infty\,.
\end{align}
A sufficient condition for this to be true is $  \xi(0)<\infty $, since
\begin{align}
\int_{k}\sum_{n=1}^{\infty} |\cos \left( kx \right)P_{n}^{\text{RPT}}(k)|=\int_{k}|\cos \left( kx \right)|\sum_{n=1}^{\infty} P_{n}^{\text{RPT}}(k)<\int_{k}\sum_{n=1}^{\infty} P_{n}^{\text{RPT}}(k)=\int_{k}P(k)=\xi(0)\,,
\end{align}
It turns out, however, that $\xi(0)$ is unbounded. This is due to the $q^{-1}$ singularity in the integrand of its expression, which is hard to overcome as $\sigma^{2}(q)\sim q^{2}$ as $q\to 0$ by construction. Since we do not believe the presence of the $|\cos \left( kx \right)|$ factor changes the divergence of the integral, a proof along these lines seems out of reach. One could also ask what the difference between SPT and RPT is in terms of the non-convergence proof for SPT. Going back to \eqref{corrfnc2} and the following discussion, we see that RPT is effectively expanding in the size of $f$. Then, interestingly, $f$ is by definition bound to be less than or equal to unity (since $\sigma^{2}(q)\geq 0$), meaning the essential singularity is just barely part of the integrand; we do not go beyond it. It is therefore not possible to conclude that perturbation theory for the integral diverges along these lines either. We plot the performance of RPT for the same initial conditions in Figure \ref{RPTplots}. Once again, we believe the result is inconclusive.


\section{1D count-in-cell PDF}\label{PDF}

Here we provide the detailed formulas and final expression for the PDF \eqref{PDFrough}, which we repeat here for convenience
\begin{align}
P_{R_{f}}(\delta_{f})d\delta_{f}=\int_{I|_{\text{satisfying Lagrangian property}}}P_{MVG}\left[\delta_{L}(q),\delta_{R_{i}}(q),\delta_{R_{i}}^{\prime}(q)\right](1-\delta_{L}(q)).
\end{align}
Let us first specify the multivariate Gaussian:
\begin{align}
P_{MVG}\left[\delta_{L}(q),\delta_{R_{i}}(q),\delta_{R_{i}}^{\prime}(q)\right]=\frac{1}{\sqrt{(2\pi)^{3} |C|}}e^{-\frac{1}2{2}\vec{v}^{T}C^{-1}\vec{v}},
\end{align}
where $\vec{v}^{T}=\left(\delta_{L}(q),\delta_{R_{i}}(q),\delta_{R_{i}}^{\prime}(q)\right)$, and $C=\langle \vec{v}\vec{v}^{T} \rangle$. More explicitly, the symmetric covariance matrix is
\begin{align}
C_{ij}=
\begin{pmatrix}
\int_{k}P_{L}(k) & \int_{k}\frac{\sin(kR)}{kR} P_{L}(k) & -\frac{1}{R}C_{12}+\frac{1}{R}\xi_{L}(R) \\
C_{12} & \int_{k}\left(\frac{2\sin(\frac{kR}{2})}{kR}\right)^{2}P_{L}(k) & -\frac{1}{R}C_{22}+\frac{1}{R}C_{12} \\
C_{13} & C_{23} & \frac{1}{R^{2}}C_{22}+\frac{1}{R^{2}}C_{11}-\frac{1}{R^{2}}C_{12} 
\end{pmatrix},
\end{align}
which is positive definite. Since the domain of integration only restricts the integrals over $\delta_{R}$ and $\delta_{R}^{\prime}$, we can perform the integral over $\delta_{L}$ already at this point, yielding
\begin{align}
\int\, d\delta_{L} &P_{MVG}\left[\delta_{L},\delta_{R_{i}},\delta_{R_{i}}^{\prime}\right](1-\delta_{L}(q))= \nonumber\\
&=P_{BiVG}\left[\delta_{R_{i}},\delta_{R_{i}}^{\prime}\right]\left(1-\frac{\delta_{R}\left(\frac{C_{12}}{C_{22}}-\frac{C_{13}C_{23}}{C_{22}C_{33}}\right)+\delta_{R}^{\prime}\left(\frac{C_{13}}{C_{33}}-\frac{C_{12}C_{23}}{C_{22}C_{33}}\right)}{1-\frac{C_{23}^{2}}{C_{22}C_{33}}}\right),
\end{align}
where $P_{BiVG}$ denotes the bivariate Gaussian distribution. Next we specify how we restrict the domain of the remaining integral. Since we are dealing with infinitesimals, we can use linear approximations everywhere. Therefore, we wish to integrate over all initial $\delta_{R}$ and $\delta_{R}^{\prime}$, such that the line $\delta_{R}+\lambda \delta_{R}^{\prime}$ crosses $\left[\{R,\bar{\delta}_{L}\},\{R+d\bar{\delta}_{L}/s,\bar{\delta}_{L}+d\bar{\delta}_{L}\}\right]$, which is a line element.This comes down to the following restriction:
\begin{align}\label{restriction}
\begin{cases}
\text{if}\quad \delta_{R}>\bar{\delta}_{L},&\, \delta_{R}^{\prime}<s\left(1+\frac{\bar{\delta}_{L}-\delta_{R}}{d\bar{\delta}_{L}}\right) \\
\text{if}\quad \delta_{R}<\bar{\delta}_{L},&\, \delta_{R}^{\prime}>s\left(1+\frac{\bar{\delta}_{L}-\delta_{R}}{d\bar{\delta}_{L}}\right) 
\end{cases}.
\end{align} 
It is in this case convenient to rewrite $\delta_{R}=\bar{\delta}_{L}+fd\bar{\delta}_{L}$, and change variables from $\delta_{R}$ to $f$, such that the PDF becomes proportional to $d\bar{\delta}_{L}$, as it should, and the above condition becomes cleaner. In fact, we can replace $\delta_{R}$ with $\bar{\delta}_{L}$, forgetting about the $fd\bar{\delta}_{L}$ correction, everywhere apart from \eqref{restriction}, since the latter is the only place where the infinitesimal drops out. One might worry that the integral over $f$ could spoil the smallness of this term. The reason we do not have to worry about this is that for large $f$, \eqref{restriction} guarantees we only integrate over $|\delta_{R}^{\prime}|\gg 1$, whose contribution to the integral is exponentially small. The remaining multivariate Gaussian integral, with the mentioned restriction, can then be written as 
\begin{align}
P_{R_{f}}(\delta_{f})d\delta_{f}=&P_{G}(\bar{\delta}_{L},C_{22})\left(\int_{0}^{\infty}\,df\int_{-\infty}^{s(1-f)}\,d\delta_{R}^{\prime}+\int_{-\infty}^{0}\,df\int_{s(1-f)}^{\infty}\,d\delta_{R}^{\prime}\right)\times \nonumber \\
&\times P_{G}\left(\delta_{R}^{\prime},\tilde{\sigma}^{2},\mu\right)\left(1-\frac{\bar{\delta}_{L}\left(\frac{C_{12}}{C_{22}}-\frac{C_{13}C_{23}}{C_{22}C_{33}}\right)+\delta_{R}^{\prime}\left(\frac{C_{13}}{C_{33}}-\frac{C_{12}C_{23}}{C_{22}C_{33}}\right)}{1-\frac{C_{23}^{2}}{C_{22}C_{33}}}\right),
\end{align}
where $\tilde{\sigma}^{2}=C_{33}(1-\kappa^{2})$, $\mu=\bar{\delta}_{L}\kappa\sqrt{\frac{C_{33}}{C_{22}}}$ is the mean of the Gaussian, and $\kappa^{2}=C_{23}^{2}/(C_{22}C_{33})$. The nontrivial integrals can now be approximated by
\begin{align}
&\left(\int_{0}^{\infty}\,df\int_{-\infty}^{s(1-f)}\,d\delta_{R}^{\prime}+\int_{-\infty}^{0}\,df\int_{s(1-f)}^{\infty}\,d\delta_{R}^{\prime}\right)\times P_{G}\left(\delta_{R}^{\prime},C_{33}(1-\kappa^{2}),\mu\right)\approx 1-\frac{\mu}{s} \nonumber \\
& \left(\int_{0}^{\infty}\,df\int_{-\infty}^{s(1-f)}\,d\delta_{R}^{\prime}+\int_{-\infty}^{0}\,df\int_{s(1-f)}^{\infty}\,d\delta_{R}^{\prime}\right)\times P_{G}\left(\delta_{R}^{\prime},C_{33}(1-\kappa^{2}),\mu\right)\times \delta_{R}^{\prime}\approx  \mu-\frac{\mu^{2}}{s}-\frac{\tilde{\sigma}^{2}}{s}.
\end{align}
These approximations are valid up to exponentially small corrections in $(s-\mu)^{2}/\tilde{\sigma}^{2}$, which lead to sub-percent corrections to the PDF. Thus, the final formula for the PDF is
\begin{align}\label{finalPDF}
P_{R_{f}}(\delta_{f})d\delta_{f}=&P_{G}(\bar{\delta}_{L},C_{22})\times \nonumber \\
&\times\left(1-\frac{\mu}{s}\right)\left(1-\frac{\bar{\delta}_{L}\left(\frac{C_{12}}{C_{22}}-\frac{C_{13}C_{23}}{C_{22}C_{33}}\right)+\left(\mu-\frac{\tilde{\sigma}^{2}}{s-\mu}\right)\left(\frac{C_{13}}{C_{33}}-\frac{C_{12}C_{23}}{C_{22}C_{33}}\right)}{1-\frac{C_{23}^{2}}{C_{22}C_{33}}}\right).
\end{align}
Using the planar collapse equations, one can write $R_{i}(R_{f},\delta_{f})$, or $R_{i}(R_{f},\delta_{L})$, and write the PDF in terms of both variables $\delta_{f}$ and $\delta_{L}$ as one pleases. The former is the observationally relevant formulation, but the latter is more convenient to compare with perturbation theory. 


\section{Non-perturbative terms in the halo model}\label{halo}

The halo model relies on the same idea we used to construct the PDF above \cite{Cooray:2002dia,Seljak:2000gq,Ma:2000ik}. The difference is that in the halo model, densities above a certain threshold collapse to form halos that have a mass dependent spatial profile (e.g. the NFW profile \cite{Navarro:1995iw}). The abundance of halos of a certain mass, which in the halo model translates into a certain initial radius $R_{i}$, is therefore given by the tail of the initial Gaussian distribution, with variance $\sigma_{R_{i}}^{2}$. This appendix does not contain original work, but simply highlights that the 1-halo term is a non-perturbative contribution to the correlation function similar to the non-perturbative errors we study in the main text.

\subsection*{Correlation function}

The 1-halo contribution to the real space two-point correlation function in the halo model is given by
\begin{align}
\xi_{1h}(r)=\frac{1}{\bar{\rho}^{2}}\int dm\, m^{2}n(m)\int dx \, \lambda_{m}(x)\lambda_{m}(x+r)
\end{align}
We care about the contribution of this term at large $r$. First observe that 
\begin{align}
\frac{m^{2}n(m)}{\bar{\rho}}\frac{dm}{m}=\sqrt{\frac{2}{\pi}}e^{-\frac{\nu^{2}}{2}}d\nu; \quad \nu=\frac{\delta_{c}}{\sigma(m)},
\end{align} 
and $\sigma(m)$ is given by the initial variance $\sigma_{R_{i}}$, and the relation $\bar{\rho}(a_{i})R_{i}=m$. The 1-halo term thus becomes
\begin{align}
\xi_{1h}(r)=\frac{1}{\bar{\rho}}\int dm\, m\sqrt{\frac{2}{\pi}}e^{-\frac{\nu^{2}}{2}}\frac{d\nu}{dm}\int dx \, \lambda_{m}(x)\lambda_{m}(x+r)\equiv \frac{1}{\bar{\rho}}\int dm\, m\sqrt{\frac{2}{\pi}}e^{-\frac{\nu^{2}}{2}}\frac{d\nu}{dm}f(r,m).
\end{align}
Now, irrespective of details of $f$ and $\sigma(m)$, it is already clear that this term is non-perturbative in the amplitude of the linear variance. Namely, derivatives of the integrand with respect to $\sigma$ are identically zero at $\sigma=0$ for all $m$. Since these derivatives are well-behaved, we can interchange derivative and the integral over $m$, and conclude that the derivatives of this contribution to the correlation function are indeed non-analytic, but obviously non-vanishing. We can try to go a bit further and estimate its behavior as a function of $r$ for large $r$. For typical halo profiles $\lambda_{m}$, larger mass halos will extend over a larger region of space. This qualitatively causes $f(r,m)$ to only have support for $m>M(r)$, where $M(r)$ is some function that computes the mass threshold. Then the integral can be written as
\begin{align}
\xi_{1h}(r)=\frac{1}{\bar{\rho}}\int_{M(r)}^{\infty} dm\, m\sqrt{\frac{2}{\pi}}e^{-\frac{\nu^{2}}{2}}\frac{d\nu}{dm}f(r,m).
\end{align}
Then, for large mass, the variance typically goes as some negative power law: $\sigma^{2}(m)\sim m^{-n}$, for $n>0$. Thus for large $r$,
\begin{align}
\xi_{1h}(r\to\infty)=\frac{1}{\bar{\rho}}\int_{M(r)}^{\infty} dm\, m\sqrt{\frac{2}{\pi}}e^{-cm^{n}}\frac{d\nu}{dm}f(r,m).
\end{align}
Due to the exponential cutoff, we can therefore very roughly estimate the large $r$ behavior of this expression as
\begin{align}
\log{\xi_{1h}(r\to\infty)}\propto -M^{n}(r),
\end{align}
which is the dominant term if this function is a power law. Generically, this is not obvious, but one can for instance naively assume the size of the halo is given by the Lagrangian radius, derived from the initial mass-scale relation, in which case the power law is linear. This therefore gives another argument for the exponentially suppressed contribution of non-perturbative terms at large distances. 

\subsection*{Power spectrum}

As is well-documented , the 1-halo term does not necessarily lead to non-analyticity in $k$ for the power spectrum \cite{Schmidt:2015gwz}. From its expression
\begin{align}
P_{1h}(k)=\frac{1}{\bar{\rho}^{2}}\int dm\, m^{2}n(m) |\lambda_{m}(k)|^{2},
\end{align}
we see that, similar to the correlation function, this expression is non-perturbative in the linear variance, since $n(m)$ is non-perturbative, while at the same time the integral is well-behaved in the $\sigma\to 0$ limit. In order to learn about the analytic properties in terms of $k$, observe that derivatives in the integrand can be interchanged with the integral over $m$. It is not a priori clear whether Fubini-Tonelli can be applied here, since we haven't specified $\lambda_{m}(k)$. We can still make progress understanding the low-$k$ limit though. First, we recall that $\lambda_{m}(k)$ must go to unity as $k\to 0$, since
\begin{align}
\lambda_{m}(k)=\int dx \, e^{ikx} \lambda_{m}(x),
\end{align}
and $\lambda_{m}$ is normalized such that its integral over space is unity. Therefore the power spectrum gets a constant contribution as $k\to 0$, which forms the classic contradiction \cite{Schmidt:2015gwz,Valageas:2010yw,Mohammed:2014lja,Seljak:2015rea} between the halo model and Peebles's argument about the $k^{4}$ scaling of non-perturbative effects from small scales at low $k$ in the power spectrum as a consequence of mass and momentum conservation \cite{Peebles:1980,Mercolli:2013bsa}. We will not go into this further at this point. Finally, we note that if $\lambda_{m}(k)$ contained some non-analytic piece, 
\begin{align}
\lambda_{m}(k)\underset{k\to 0}{=}1+\text{analytic terms }+ \# e^{-\frac{1}{k^{2}}},
\end{align}
it would not be captured by any Taylor expansion in small $k$ of the power spectrum. We leave it to future work to see whether such terms exist, and what their relation to Peebles's argument is.  

\bibliographystyle{utphys}
\bibliography{Paper}
%
%
%
%
%
%

\end{document}